\documentclass[aps,prd,twocolumn,superscriptaddress,amsmath,amssymb]{revtex4}

\usepackage{etoolbox}
\newbool{useHFLAV}
\setbool{useHFLAV}{true}

\usepackage{environ}
\NewEnviron{comment}{}

\newcommand\coff{\RenewEnviron{comment}{\BODY}}
\newcommand{\cmt}[1]{\begin{comment}{#1}\end{comment}\xspace}
\coff

\usepackage{graphicx} 
\usepackage{epstopdf} 
\usepackage{dcolumn} 
\usepackage{xcolor} 
\usepackage{amsmath,amssymb} 
\usepackage[hidelinks]{hyperref} 
\usepackage[utf8]{inputenc} 
\usepackage[english]{babel} 
\usepackage[displaymath, mathlines]{lineno} 
\usepackage{blindtext} 
\usepackage{ifthen} 
\usepackage{orcidlink} 
\usepackage{multirow}

\graphicspath{{figures/}} 

\newboolean{articletitles}
\setboolean{articletitles}{true} 

\input{belle2-symbols}

\ifbool{useHFLAV}{

    \def\acpqqs{\cmt{\ensuremath{\ACP = 0.01\pm 0.14}}}
    \def\acpnrkwa{\cmt{\ensuremath{\ACP = 0.06\pm0.08}}}
}{

    \def\acpqqs{\cmt{\ensuremath{\ACP = -0.01\pm 0.14}}}
    \def\acpnrkwa{\cmt{\ensuremath{\ACP = -0.06\pm0.08}}}
}

\begin{document}


\title{Measurement of $\boldsymbol{\CP}$ asymmetries in $\boldsymbol{\phiks}$ decays with \belletwo }
  \author{I.~Adachi\,\orcidlink{0000-0003-2287-0173}} 
  \author{K.~Adamczyk\,\orcidlink{0000-0001-6208-0876}} 
  \author{L.~Aggarwal\,\orcidlink{0000-0002-0909-7537}} 
  \author{H.~Ahmed\,\orcidlink{0000-0003-3976-7498}} 
  \author{H.~Aihara\,\orcidlink{0000-0002-1907-5964}} 
  \author{N.~Akopov\,\orcidlink{0000-0002-4425-2096}} 
  \author{A.~Aloisio\,\orcidlink{0000-0002-3883-6693}} 
  \author{N.~Anh~Ky\,\orcidlink{0000-0003-0471-197X}} 
  \author{D.~M.~Asner\,\orcidlink{0000-0002-1586-5790}} 
  \author{H.~Atmacan\,\orcidlink{0000-0003-2435-501X}} 
% ...
\collaboration{The Belle II Collaboration}

\begin{abstract}
We present a measurement of time-dependent rate asymmetries in \phiks decays to search for non-standard-model physics in \qqs transitions.
The data sample is collected with the \belletwo detector at the SuperKEKB asymmetric-energy \epem collider in 2019--2022 and contains \nbb bottom-antibottom mesons from \FourS resonance decays.
We reconstruct \ensuremath{162 \pm 17}\xspace signal events and extract the charge-parity (\CP) violating parameters from a fit to the distribution of the proper-decay-time difference of the two \B mesons.
The measured direct and mixing-induced \CP asymmetries are $\ACP=\cmt{\ensuremath{0.31 \pm 0.20 \pm0.05}}\xspace$ and $\SCP=\ensuremath{0.54 \pm 0.26 ^{+0.06}_{-0.08}}\xspace$, respectively, where the first uncertainties are statistical and the second are systematic.
The results are compatible with the \CP asymmetries observed in \ccs transitions.

\end{abstract}

\maketitle

\section{Introduction}
Measurements of \CP asymmetries in loop-suppressed \B meson decays are sensitive probes of physics beyond the standard model (SM).
In particular, gluonic-penguin \qqs modes, such as  \phiks, are sensitive to interfering non-SM amplitudes that carry additional weak-interaction phases.
The SM reference is the mixing-induced \CP asymmetry $\SCP \equiv \sin2\phi_1$ observed in tree-level \ccs transitions, where $\phi_1$ (or $\beta$) equals $\arg(-\Vcd\Vcb^*/\Vtd\Vtb^*)$ and \Vij are Cabibbo-Kobayashi-Maskawa (CKM) quark-mixing matrix elements~\cite{Cabibbo:1963yz,Kobayashi:1973fv}.
The deviation from the value of \SCP observed in \ccs transitions, $\SCP=0.699\pm0.017$~\cite{HeavyFlavorAveragingGroup:2022wzx}, is the key observable. 
For \phiks decays, such a deviation is at most $0.02\pm0.01$ within the SM while the direct \CP asymmetry \ACP is expected to be zero~\cite{Beneke:2005pu}. 
The current world-average values for \phiks are $\SCP = 0.74 ^{+0.11}_{-0.13}$ and \acpqqs~\cite{HeavyFlavorAveragingGroup:2022wzx}.
Therefore, experimental knowledge must be improved. 
We present a measurement of \SCP and \ACP in the sample of electron-positron collisions collected by the Belle II experiment in 2019–2022~\cmt{\cite{cp-coeffs}}.

At \B-factories, \BBbar events are produced from the decay of an \FourS resonance, where \B indicates a \Bp or \Bz meson.
We denote pairs of neutral \B mesons as $\bsig\btag$, where \bsig decays into a \CP-eigenstate at time \tcp, and \btag decays into a flavor-specific final state at time \ttag.
For quantum-correlated \B-meson pairs, the flavor of \bsig is opposite to that of \btag at the instant when the \btag decays.
The probability to observe a \btag meson with flavor $q$ ($q=+1$ for \Bz and $q=-1$ for \Bzb) and a proper-time difference $\Delta t \equiv \tcp - \ttag$ between the \bsig and \btag decays is
\begin{linenomath}
\begin{align}
\begin{split}
\mathcal{P}(\dt,q)= \frac{e^{-|\dt|/\taud}}{4\taud}  \Big\{ 1 + q\big[ &  \SCP \sin(\dmd \dt)\\
\ifbool{useHFLAV}{
    - & \ACP \cos (\dmd \dt) \big] \Big\},
}{
    + & \ACP \cos (\dmd \dt) \big] \Big\},
}
\end{split}
\label{eq:dt_theo}
\end{align}
\end{linenomath}
where \taud and \dmd are the \Bz lifetime and $\Bz-\Bzb$ mixing frequency, respectively~\cite{Workman:2022ynf}.

We reconstruct \phiks decays in a sample of energy-asymmetric \epem collisions at the \FourS resonance provided by SuperKEKB and collected with the \belletwo detector.
The sample corresponds to \lumi and contains \nbb \BBbar events.
We fully reconstruct \bsig in the $\phi\KS$ final state using the intermediate decays \phikk and \kspipi, while we only determine the position of the \btag decay.
The flavor of the \btag meson is inferred from the properties of all charged particles in the event not belonging to \bsig~\cite{Belle-II:2021zvj}.
In order to extract the \CP asymmetries, we model the distributions of signal \bsig and backgrounds in \dt and other discriminating variables, and then perform a likelihood fit. 
The last measurements, by the \belle and \babar experiments, used time-dependent Dalitz-plot analyses~\cite{PhysRevD.82.073011,PhysRevD.85.112010}.
This method models the interferences among the intermediate resonant and nonresonant amplitudes contributing to \nonres decays, thereby providing the best sensitivity on $\phi_1$.
Due to the small dataset size, which may induce multiple solutions in the Dalitz-plot approach, we perform a quasi-two-body analysis by restricting the sample to candidates reconstructed in a narrow region around the $\phi$ mass. 
This strategy offers the advantage of a simpler analysis, albeit with a reduced statistical sensitivity.
We use the knowledge from the previous Dalitz-plot analyses to estimate the effect of neglecting the interferences.
We test our analysis on the \CP-conserving \phikp decay, which has similar backgrounds and vertex resolution.
Charge-conjugated modes are included throughout the paper.

\section{Experimental setup}
\label{sec:experiment}
The \belletwo detector~\cite{Abe:2010gxa} operates at the SuperKEKB accelerator at KEK, which collides 7~\gev electrons with 4~\gev positrons.
The detector is designed to reconstruct the decays of heavy-flavor mesons and $\tau$ leptons.
It consists of several subsystems arranged cylindrically around the interaction point (IP).
The innermost part of the detector is equipped with a two-layer silicon-pixel detector (PXD), surrounded by a four-layer double-sided silicon-strip detector (SVD)~\cite{Belle-IISVD:2022upf}.
Together, they provide information about charged-particle trajectories (tracks) and decay-vertex positions.
Of the outer PXD layer, only one-sixth is installed for the data used in this work.
The momenta and electric charges of charged particles are determined with a 56-layer central drift-chamber (CDC).
Charged-hadron identification (PID) is provided by a time-of-propagation counter and an aerogel ring-imaging Cherenkov counter, located in the central and forward regions outside the CDC, respectively.
The CDC provides additional PID information through the measurement of specific ionization.
Photons are identified and electrons are reconstructed by an electromagnetic calorimeter made of CsI(Tl) crystals, covering the region outside of the PID detectors.
The tracking and PID subsystems, and the calorimeter, are surrounded by a superconducting solenoid, providing an axial magnetic field of 1.5~T.
The central axis of the solenoid defines the $z$ axis of the laboratory frame, pointing approximately in the direction of the electron beam.
Outside of the magnet lies the muon and \KL identification system, which consists of iron plates interspersed with resistive-plate chambers and plastic scintillators.

We use simulated events to model signal and background distributions, study the detector response, and test the analysis.
Quark-antiquark pairs from \epem collisions, and hadron decays, are simulated using \kkmc~\cite{Jadach:1999vf} with \pythia~\cite{Sjostrand:2014zea}, and \evtgen~\cite{Lange:2001uf}, respectively.
The detector response and \KS decays are simulated using \geant~\cite{Agostinelli:2002hh}.
Collision data and simulated samples are processed using the \belletwo analysis software~\cite{Kuhr:2018lps,basf2-zenodo}.

\section{Event reconstruction}
Events containing a \BBbar pair are selected online by a trigger based on the track multiplicity and total energy deposited in the calorimeter.
We reconstruct \phiks decays using \phikk and \kspipi decays, in which the four tracks are reconstructed using information from the PXD, SVD, and CDC~\cite{Bertacchi:2020eez}. 
All tracks are required to have polar angle $\theta$ within the CDC acceptance ($17^\circ<\theta<150^\circ$).
Tracks used to form \Phi candidates are required to have a distance of closest approach to the IP less than $2.0$~\cm along the $z$ axis and less than $0.5$~\cm in the transverse plane to reduce contamination of tracks not generated in the collision.

Kaon and pion mass hypotheses are assigned to tracks based on information provided by the PID subsystems.
The \Phi candidates are formed by combining $\Kp\Km$ pairs consistent with originating from the IP and having invariant mass within $[0.99,1.09]$~\gevcc, where the average $\phi$ mass resolution is approximately 3~\mevcc.
The \KS candidates are formed by combining two oppositely charged particles, assumed to be pions, and requiring their invariant mass to be within $[0.480,0.515]$~\gevcc, where the average \KS mass resolution is approximately 2~\mevcc.
In order to suppress combinatorial background from misreconstructed \KS, we require \KS candidates to have a displacement of at least 0.05~\cm from the $\phi$ decay vertex, where the average \KS flight distance is 10~\cm.

The beam-energy constrained mass \mbc and energy difference \deltae are computed for each \phiks candidate as $\mbc \equiv \sqrt{(\ebeam/c^2)^2-(|\pstar|/c)^2}$ and $\deltae \equiv \estar - \ebeam$, where \ebeam is the beam energy, and \estar and \pstar are the energy and momentum of the \bsig candidate, respectively, all calculated in the center-of-mass (c.m.) frame.
Signal \bsig candidates peak at the known \Bz mass~\cite{Workman:2022ynf} and zero in \mbc and \deltae, respectively, while continuum is distributed more uniformly.
Only candidates satisfying $\mbc>5.2~\gevcc$ and $|\deltae|<0.2~\gev$ are retained for further analysis.

The \phiks decay vertex is determined using the \treefit algorithm~\cite{HULSBERGEN2005566, Krohn:2019dlq}. 
In addition, the \bsig candidate is constrained to point back to the IP.
The \btag decay vertex is reconstructed using the remaining tracks in the event.
Each track is required to have at least one measurement point in the SVD and CDC subdetectors and correspond to a total momentum greater than 50~\mevc.
The \btag decay-vertex position is fitted using the \rave algorithm~\cite{Waltenberger_2008}, which allows for weighting the contributions from tracks that are displaced from the \btag decay vertex, and thereby suppressing biases from secondary charm decays.
The decay-vertex position is determined by constraining the \btag direction, as determined from its decay vertex and the IP, to be collinear with its momentum vector~\cite{btube-conf}.

We estimate the proper-time difference using the longitudinal decay-vertex positions, \lcp and \ltag, of the \bsig and \btag mesons, respectively, as
\begin{equation}
\dt \approx \frac{\lcp-\ltag}{\betagamma\gamma^* c},
\label{eq:dt-approx}
\end{equation}
where $\betagamma=\boost$ is the \FourS Lorentz boost and $\gamma^*= 1.002$ is the Lorentz factor of the \B mesons in the c.m. frame.
The average distance between the \bsig and \btag vertices is approximately 100~{\textmu}m along the $z$ axis. 
The \B-decay vertex resolution along the $z$ axis is approximately 35~{\textmu}m for simulated \phiks decays.
We apply loose \chisq probability requirements to both the \bsig and \btag vertices. 
Events having a \dt uncertainty \dterr greater than $2.0$~\ps, where the average value is approximately 0.5~\ps, are not included in the analysis, as they constitute less than 2\% of the signal events and do not contribute to the determination of \SCP.

The dominant sources of background come from continuum $\epem\to\qqbar$ events, where $q$ indicates a \uquark, \dquark, \cquark, or \squark quark.
A boosted-decision-tree (BDT) classifier is trained on simulated samples to combine several topological variables that provide separation between continuum and signal events~\cite{DBLP:journals/corr/ChenG16}.
The variables included in the BDT are the following, in order of decreasing discriminating power: the cosine of the angle between the thrust axes of \bsig and \btag~\cite{BaBar:2014omp}, the modified Fox-Wolfram moments introduced in Ref.~\cite{PhysRevLett.91.261801}, the thrust of \btag \cite{Brandt:1964sa, Farhi:1977sg}, the ratio of the zeroth to the first Fox-Wolfram moment~\cite{PhysRevLett.41.1581}, and the harmonic moments calculated with respect to the thrust axis.
We impose a minimum requirement on the output of the BDT, \obdt, that retains more than $95\%$ of the signal, while rejecting more than 55\% of the continuum events.
The transformed output of the classifier, defined as $\ocs=\log[(\obdt-\obdt^{\text{min}})/(\obdt^{\text{max}}-\obdt)]$, where $\obdt^{\text{min}}$ and $\obdt^{\text{max}}$ are the minimum and maximum values of the selected events, is included in the fit. 
The signal and remaining background events are approximately Gaussian-distributed in this variable and are therefore simple to model.

An additional requirement $|\deltae|<50~\mev$ further suppresses continuum and misreconstructed \phikstar decays.
To reduce the contamination from nonresonant \nonres decays and other modes leading to the same final state, events are required to satisfy $|\mkk-m_{\phi}|<10\mevcc$, where $m_{\phi}$ is the known $\phi$ meson mass~\cite{Workman:2022ynf}.

The same event reconstruction is applied on \phikp decays, except for the \KS selection, which is replaced by a \Kp track with a stringent PID requirement.
This is more than $90\%$ efficient on the signal, while rejecting around $30\%$ of misidentified charged particles.
We achieve a total signal reconstruction efficiency of $33$\% for \phiks and $40$\% for \phikp.

Events with multiple candidates account for approximately 6\% of the data.
We keep the candidate with the highest \bsig vertex \chisq probability. 
The criterion retains the correct signal candidate 67\% of the times using simulated events.
We check that the candidate selection does not bias the \dt distribution by comparing the results of lifetime fits to the \Bz and \Bp samples with known values~\cite{Workman:2022ynf}.

\section{Time-dependent \CP-asymmetry fit}
\begin{figure*}[htpb]
\centering
\includegraphics[width=0.3\textwidth]{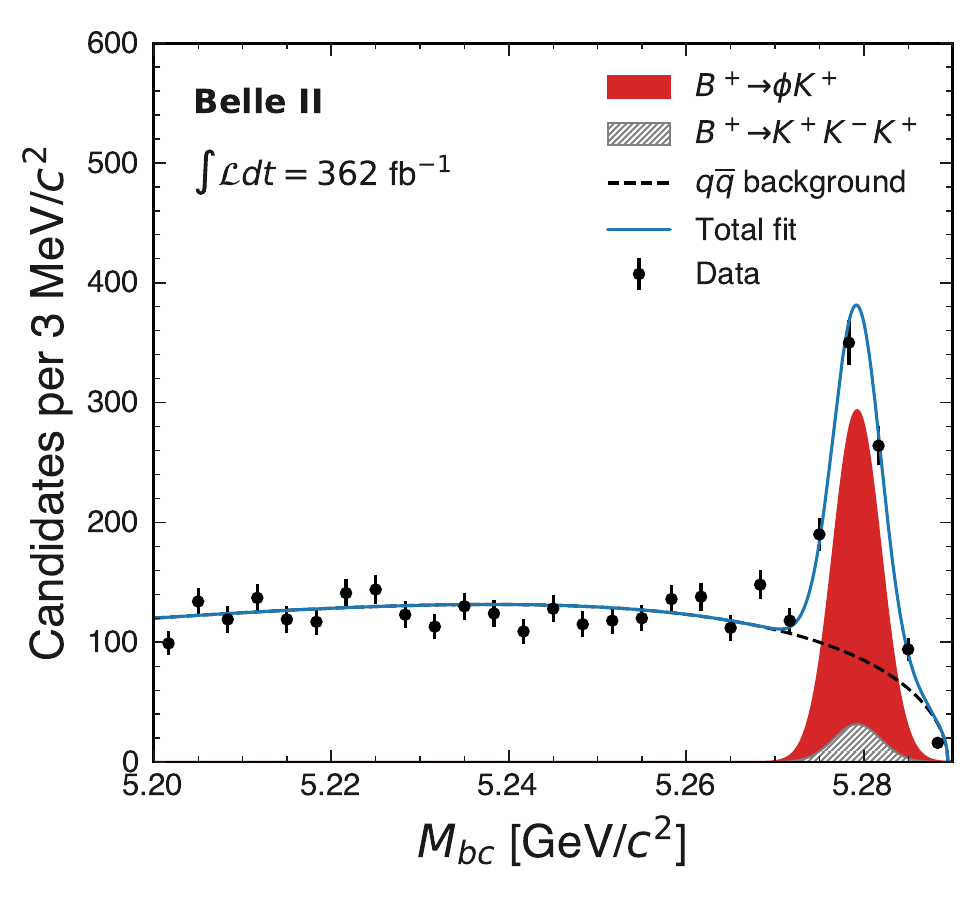}
\includegraphics[width=0.3\textwidth]{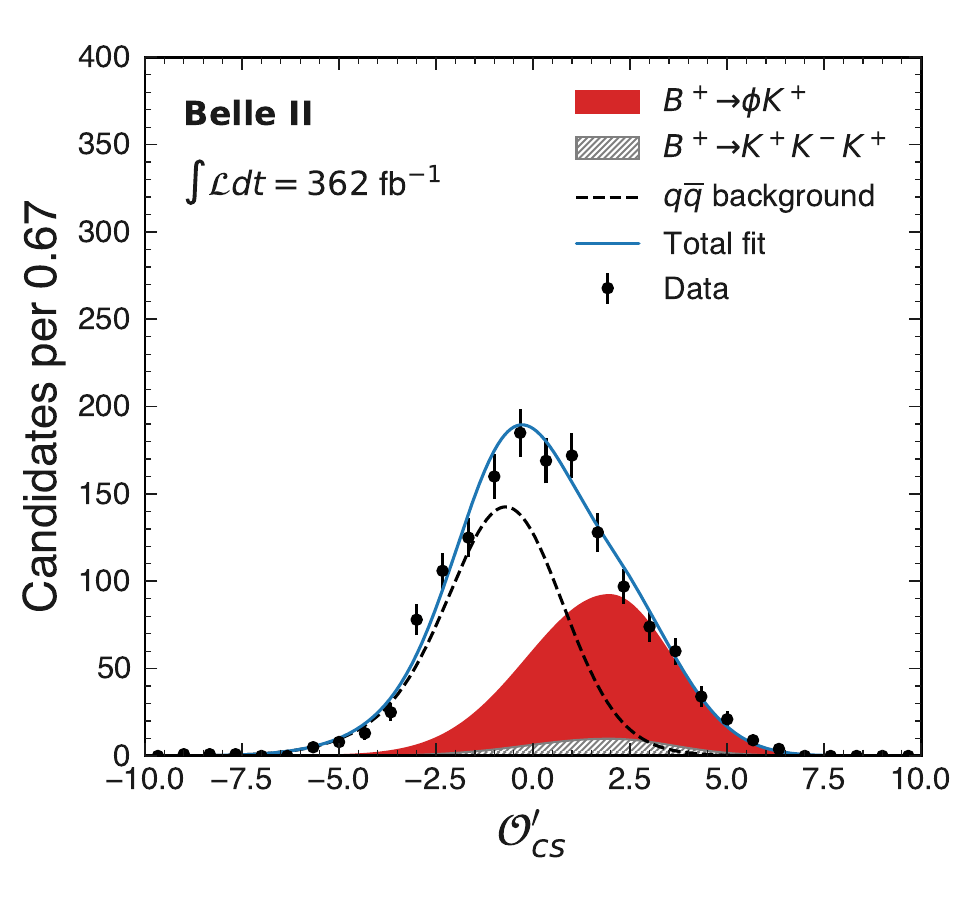}
\includegraphics[width=0.3\textwidth]{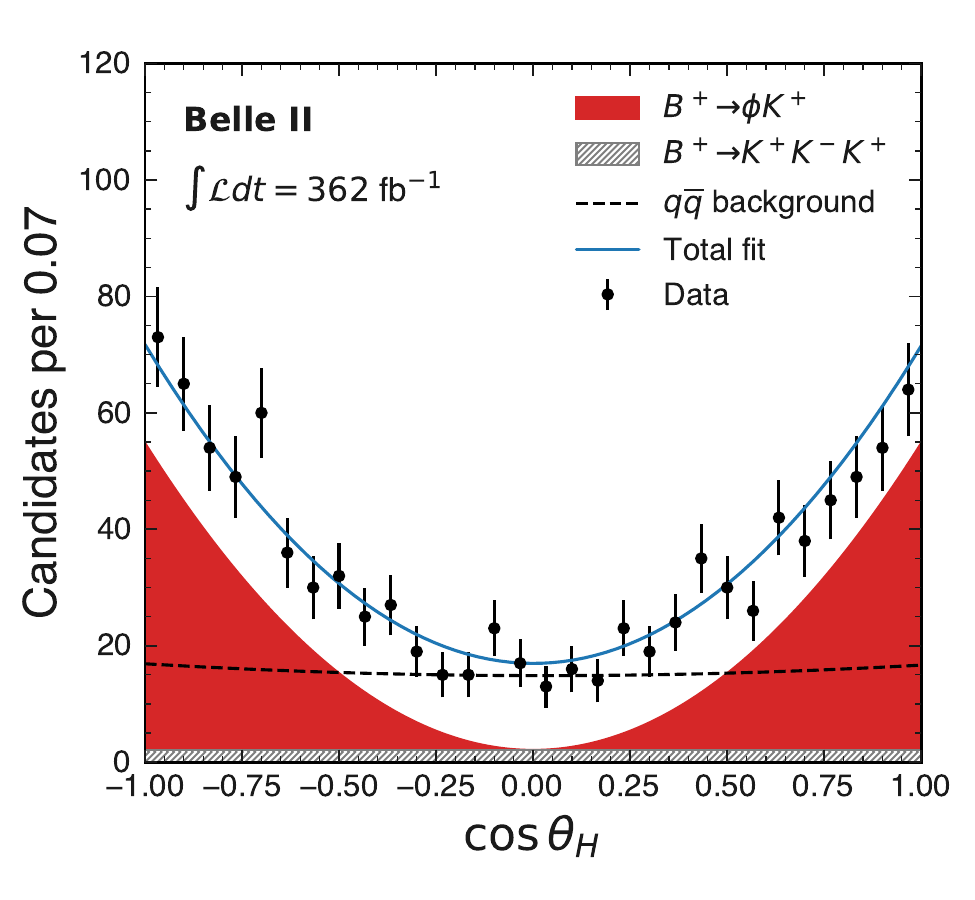}
\includegraphics[width=0.3\textwidth]{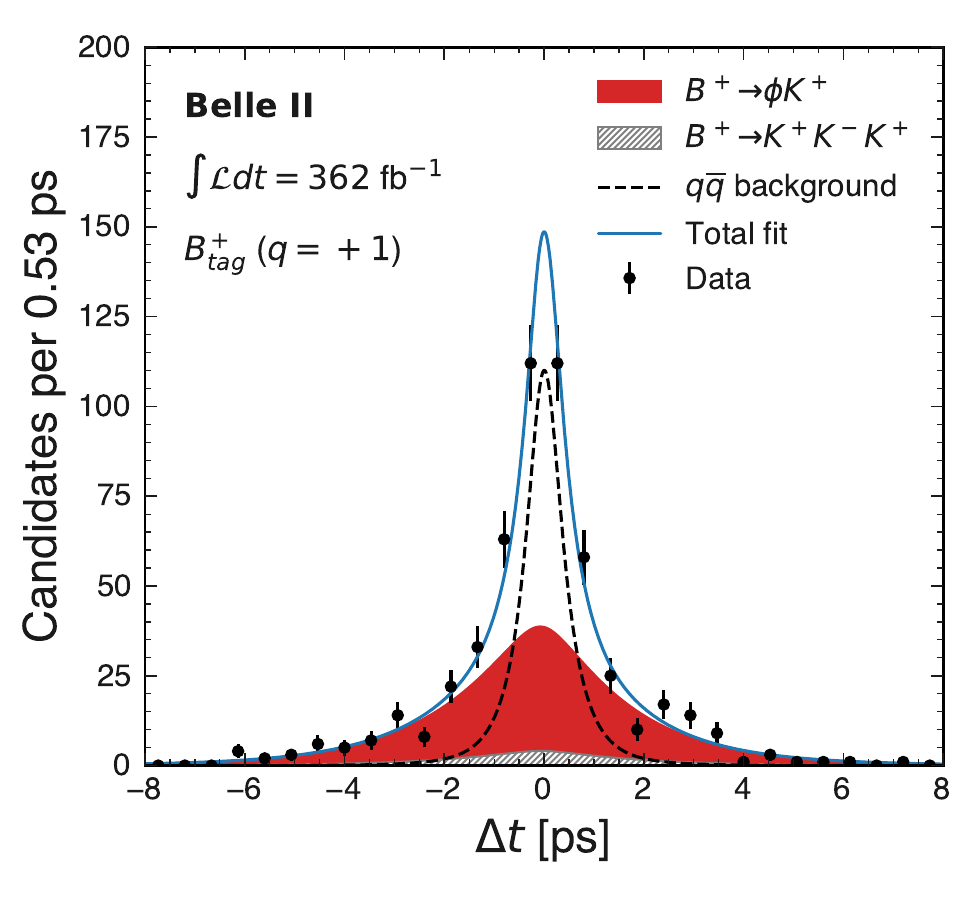} 
\includegraphics[width=0.3\textwidth]{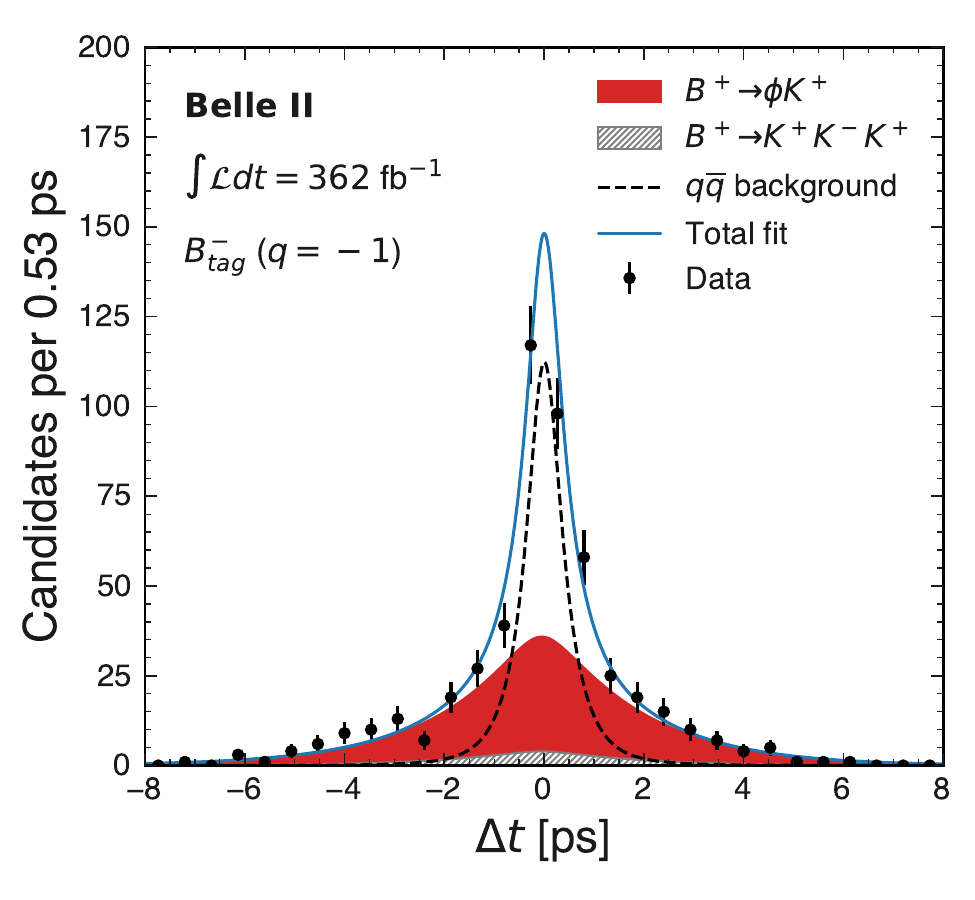} 
\caption{
Distributions of (top left) \mbc, (top center) \ocs, (top right) \cosh, (bottom left) \dt for \Bp-tagged and (bottom right) \dt for \Bm-tagged \phikp candidates (data points) with fits overlaid (curves and stacked shaded areas).
The \mbc distribution is displayed for candidates with $\ocs>-1$ and the \ocs distribution is displayed for candidates with $\mbc>5.27$~\gevcc.
The \cosh and \dt distributions are displayed for candidates with $\ocs>-1$ and $\mbc>5.27$~\gevcc.
}
\label{fig:yields-fit-kp}
\end{figure*}
\begin{figure*}[htpb]
\centering
\includegraphics[width=0.3\textwidth]{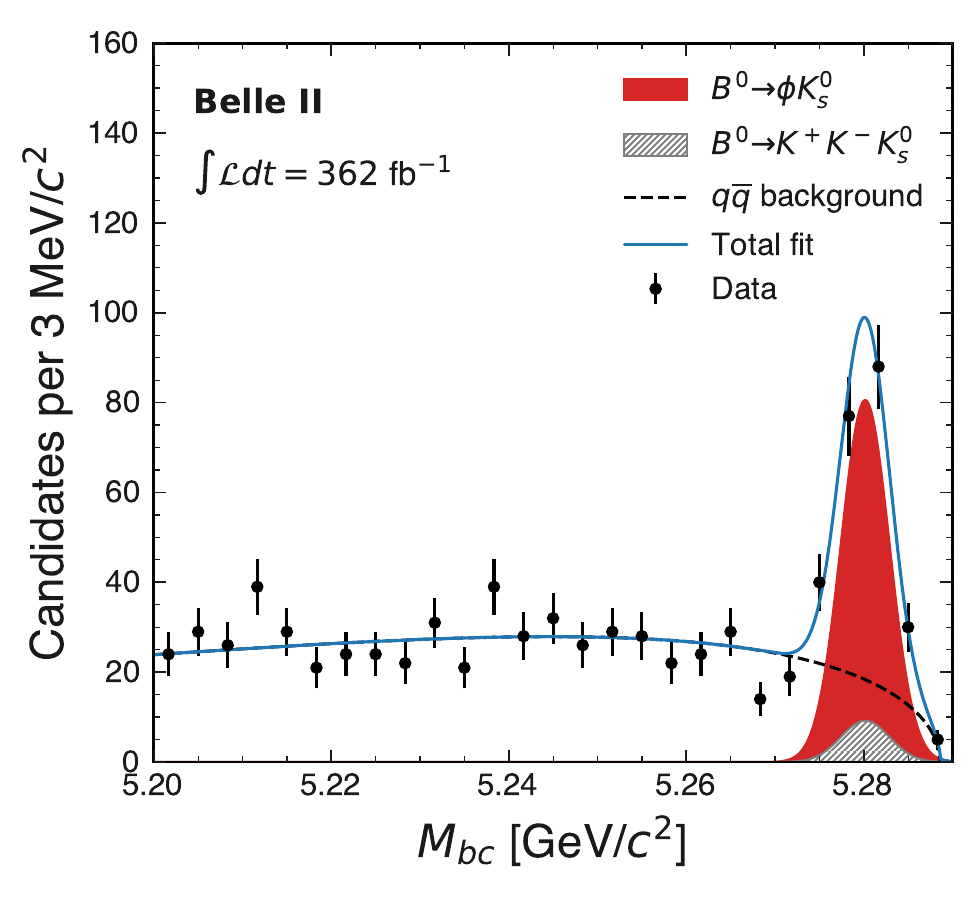}
\includegraphics[width=0.3\textwidth]{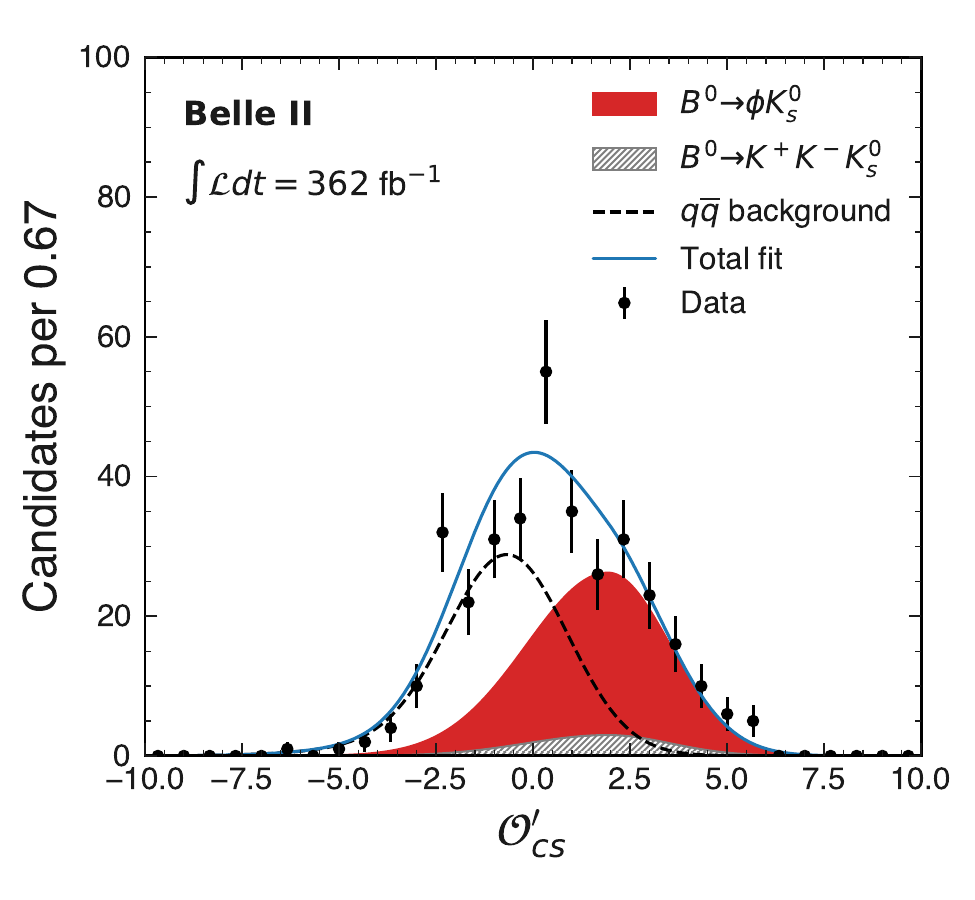}
\includegraphics[width=0.3\textwidth]{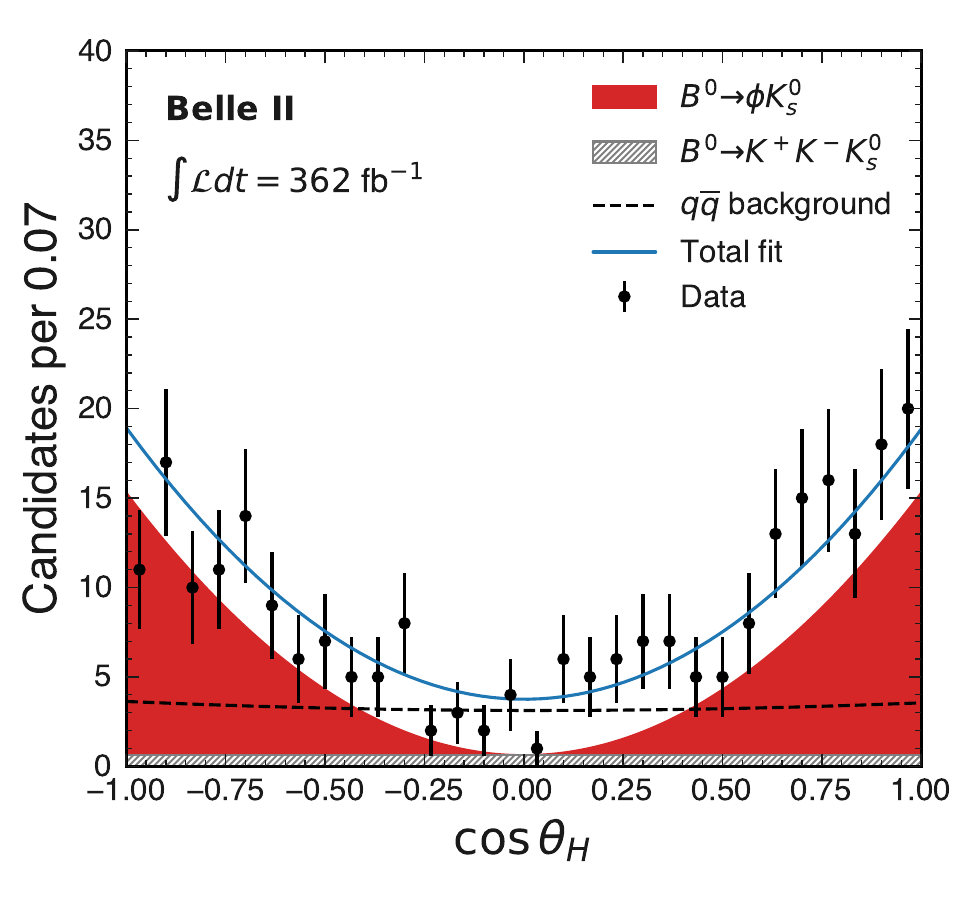}
\includegraphics[width=0.3\textwidth]{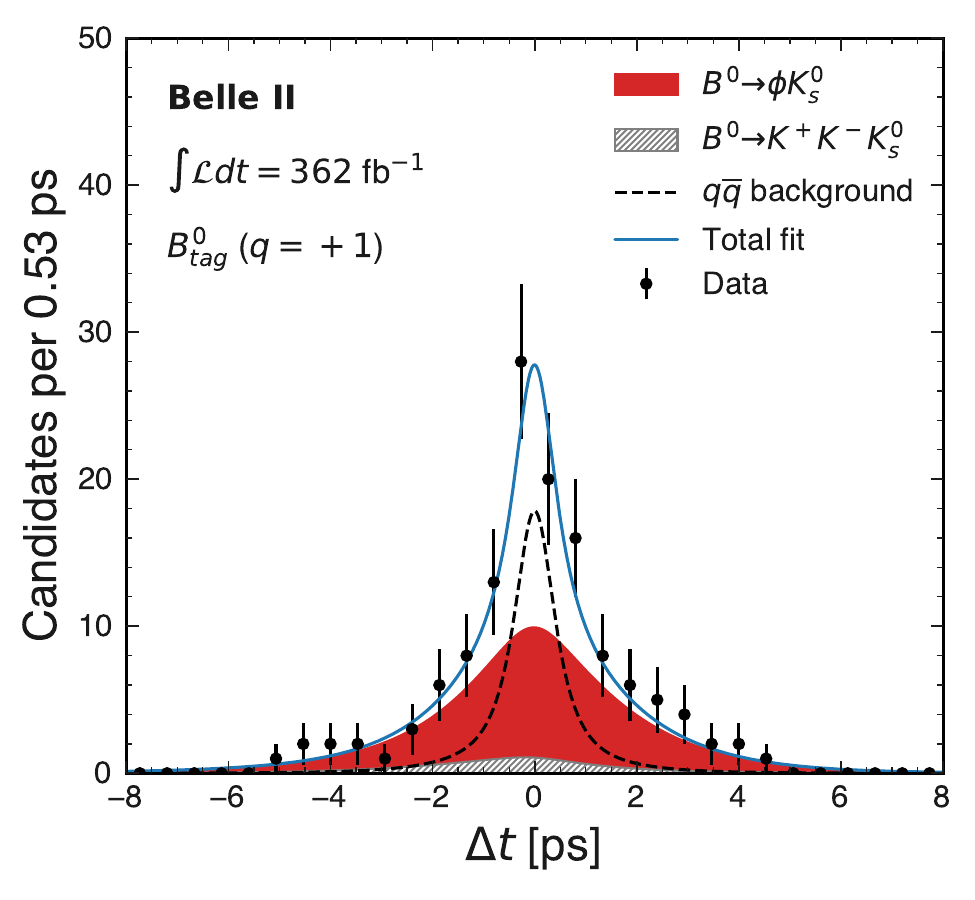} 
\includegraphics[width=0.3\textwidth]{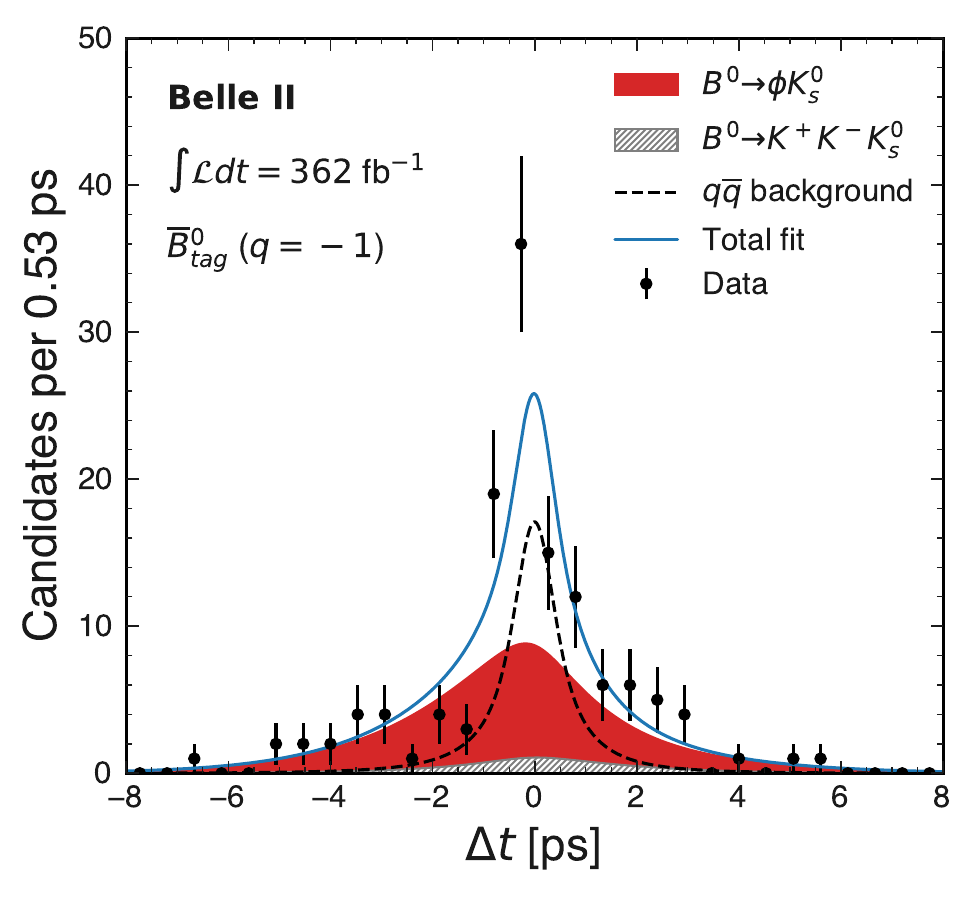} 
\caption{
Distributions of (top left) \mbc, (top center) \ocs, (top right) \cosh, (bottom left) \dt for \Bz-tagged and (bottom right) \dt for \Bzb-tagged \phiks candidates (data points) with fits overlaid (curves and stacked shaded areas).
The \mbc distribution is displayed for candidates with $\ocs>-1$ and the \ocs distribution is displayed for candidates with $\mbc>5.27$~\gevcc.
The \cosh and \dt distributions are displayed for candidates with $\ocs>-1$ and $\mbc>5.27$~\gevcc.
}
\label{fig:yields-fit-ks}
\end{figure*}

The distributions of signal and backgrounds are described in a likelihood fit to extract the \CP asymmetries.
We consider the following contributions to the sample composition: signal \phiks events, nonresonant \nonres background, and continuum background.
Additional \BBbar background events are treated as a source of systematic uncertainty, as they are estimated to be at most 2\% of the signal yield, according to simulation.
Low-multiplicity events contribute at less than the level of the \BBbar backgrounds in the simulation, and are distributed like continuum in the variables used in the fit, so they are treated as part of the continuum background.
We model the distributions of signal and background events in the \mbc, \ocs, \cosh, and \dt variables. 
The \mbc and \ocs variables provide discrimination between signal and continuum background.
The helicity angle $\theta_H$, defined as the angle between the momentum of the \Bz and that of the positively charged kaon in the $\phi$ rest frame, is used to distinguish between signal and nonresonant components.
The \dt variable and tag-flavor $q$ provide access to the time-dependent \CP asymmetries.
In addition, we use \dterr as a conditional observable to model the per-event resolution.

We extract the \CP asymmetries using an extended maximum-likelihood fit to the unbinned distributions of the discriminating variables.
The total probability density function (PDF) is given by the product of the four one-dimensional PDFs, since the dependences among the fit observables are negligible.
We model the \mbc distribution using an ARGUS function~\cite{ALBRECHT1990278} for continuum and a Gaussian function with shared parameters for the \phiks and \nonres components.
The continuum shape is fixed from a fit to the $|\deltae|>0.1~$\gev sideband, while the signal-shape parameters are determined by the fit.
We check that the continuum shapes are not biased by \phikstz, \phikstp, and other \Bz and \Bp decay modes, contributing in total to less than 1\% of the events in the \deltae sideband.
The \ocs distribution is modeled using the sum of two Gaussian functions with a common mean and constrained proportions for continuum, and a Gaussian function with asymmetric widths and shared parameters for the \phiks and \nonres components.
The \ocs shape-parameters are determined from events in the \deltae sideband for continuum, and using simulated events for signal.
The \cosh distribution of continuum is modeled with a second-order polynomial determined from \deltae sideband events.
We verify using simulated samples that the \phiks and \nonres components follow a $\cos^2\theta_{H}$ and a uniform distribution, respectively, as expected from angular momentum conservation, and the detector acceptance does not affect their shapes.

The \btag flavor is identified using a category-based \B-flavor tagging algorithm from the particles in the event that are not associated with the \bsig candidate~\cite{Belle-II:2021zvj}.
The tagging algorithm provides for each \btag candidate a flavor ($q$) and the tag-quality $r=1-2w$.
The latter is a function of the wrong-tag probability \wtag and ranges from $r=0$ for no discrimination power to $r=1$ for unambiguous flavor assignment.
Taking into account the effect of imperfect flavor assignment, Eq.~\eqref{eq:dt_theo} becomes
\begin{widetext}
\begin{linenomath}
\begin{equation}
\begin{split}
\mathcal{P}(\dt, q)  = \frac{e^{-|\dt|/\taud}}{4\taud} \Big\{ 1-\qtag \dwtag + \qtag \mutag (1-2\wtag)
+ \big[\qtag(1-2\wtag)&+\mutag(1-q\dwtag)\big] \\
\times & \big[  \SCP \sin(\dmd\dt) 
\ifbool{useHFLAV}{
    - \ACP \cos(\dmd\dt) \big] \Big\},
}{
    + \ACP \cos(\dmd\dt) \big] \Big\},
}
\end{split}
\label{eq:dt_tag}
\end{equation}
\end{linenomath}
\end{widetext}
where \dwtag is the wrong-tag probability difference between events tagged as \Bz and \Bzb, and \mutag is the tagging-efficiency-asymmetry between \Bz and \Bzb.

The effect of finite \dt resolution is taken into account by modifying Eq.~\eqref{eq:dt_tag} as follows:
\begin{linenomath}
\begin{equation}
\mathcal{F}(\dt,q | \dterr) = \int \mathcal{P}(\dtp, \qtag) \dtres d\dtp,
\label{eq:dt_res}
\end{equation}
\end{linenomath}
where $\mathcal{R}$ is the resolution function, conditional on the per-event \dt uncertainty $\dterr$.
Its parametrization, as determined in \dpi decays~\cite{Belle-II:2023bps}, consists of the sum of three components,
\begin{linenomath}
\begin{align}
\begin{split}
\mathcal{R}(\delta t | & \sigma_{\dt}) = (1-f_t-f_{\text{OL}})G(\delta t|m_G\sigma_{\dt}, s_G \sigma_{\dt}) \\
+ &f_t(\sigma_{\dt}) R_t(\delta t| m_t \sigma_{\dt},  s_t \sigma_{\dt}, k/\sigma_{\dt}, f_{>}, f_{<}) \\
+ &f_{\text{OL}} G(\delta t | 0,\sigma_{0}),
\end{split}
\end{align}
\end{linenomath}
where $\delta t$ is the difference between the observed and the true \dt. 
The first component is described by a Gaussian function with mean $m_G$ and width $s_G$ scaled by $\sigma_{\Delta t}$, which accounts for the core of the distribution.
The second component $R_t$ is the sum of a Gaussian function and the convolution of a Gaussian with two oppositely sided exponential functions,
\begin{linenomath}
\begin{align}
\begin{split}
R_t(x| \mu,  & \sigma, k, f_{>}, f_{<}) =  (1-f_{<}-f_{>}) G(x|\mu,\sigma) \\
+ & f_{<} G(x|\mu,\sigma) \otimes k \exp_{<}(kx) \\
+ & f_{>} G(x|\mu,\sigma) \otimes k \exp_{>}(-kx),
\end{split}
\end{align}
\end{linenomath}
where $\exp_{>}(kx)=\exp(kx)$ if $x>0$ or zero otherwise, and similarly for $\exp_{<}(kx)$.
The exponential tails arise from intermediate displaced charm-hadron vertices from the \btag decay.
The fraction $f_t$ is zero at low values of $\sigma_{\dt}$ and steeply reaches a plateau of 0.2 at $\sigma_{\dt}=0.25~\ps$.
The third component, which accounts for outlier events contributing with a fraction of less than 1\%, is modeled with a Gaussian function having a large width $\sigma_0$ of 200~\ps.
The effect on the resolution function of the small momentum of the \Bz in the \FourS frame is taken into account as a systematic uncertainty.

We divide our sample into seven intervals (bins) of the tag-quality variable $r$, with boundaries $(0.0,0.1,0.25,0.45,0.6,0.725,0.875,1.0)$, to gain statistical sensitivity from events with different wrong-tag fractions.
The response of the tagging algorithm and detector \dt resolution is calibrated from a simultaneous fit of $w$, $\Delta w$, \mutag, and resolution-function parameters in the seven $r$-bins, using flavor-specific \dpi decays~\cite{Belle-II:2023nmj}.
The effective flavor tagging efficiency, defined as $\sum_{i}\varepsilon_i(1-2w_i)^2$, where $\varepsilon_i$ is the fraction of events associated with a tag decision and $w_i$ is the wrong-tag probability in the $i$th $r$ bin, is $(31.69\pm0.35)\%$, where the uncertainty is statistical.
We verify in simulation the compatibility of the flavor tagging and resolution function between the calibration and signal decay modes.
We use the flavor-tagging parameters obtained from \dzpip decays to calibrate the flavor tagger and resolution function in the \phikp control channel.

The \dt distribution of the continuum background is modeled using events from the \deltae sideband and allowing for an asymmetry in the yields of oppositely tagged events.
A double Gaussian parametrization, with means and widths scaled by $\sigma_{\dt}$, describes the data accurately.
The \dt distribution of the \nonres background is parametrized using the same detector response as for signal.
Its \CP asymmetries are fixed to the known values~\cite{HeavyFlavorAveragingGroup:2022wzx}.

The nominal fits to the control and signal samples determine the continuum yields and the sum of the resonant and nonresonant yields in the seven $r$-bins. 
We also determine the fraction of the resonant yields with respect to the sum of the resonant and nonresonant yields directly in the data. 
In addition, the mean and width of the Gaussian function describing the resonant and nonresonant components in \mbc and the asymmetry in the normalization of oppositely tagged continuum-background events are determined by the fit.
Finally, the fit determines the \CP asymmetries, for a total of 20 free parameters.

\begin{figure*}[htpb]
\centering
\includegraphics[width=\columnwidth]{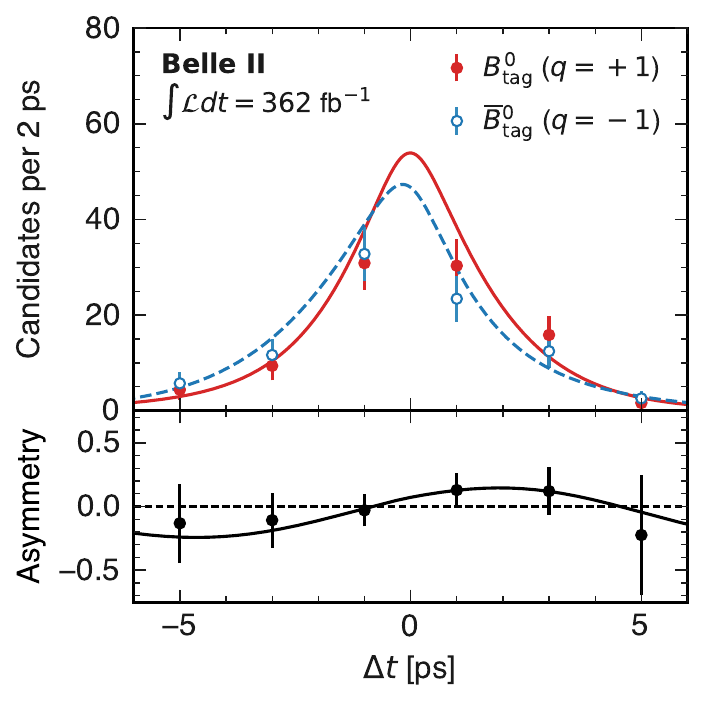}
\includegraphics[width=\columnwidth]{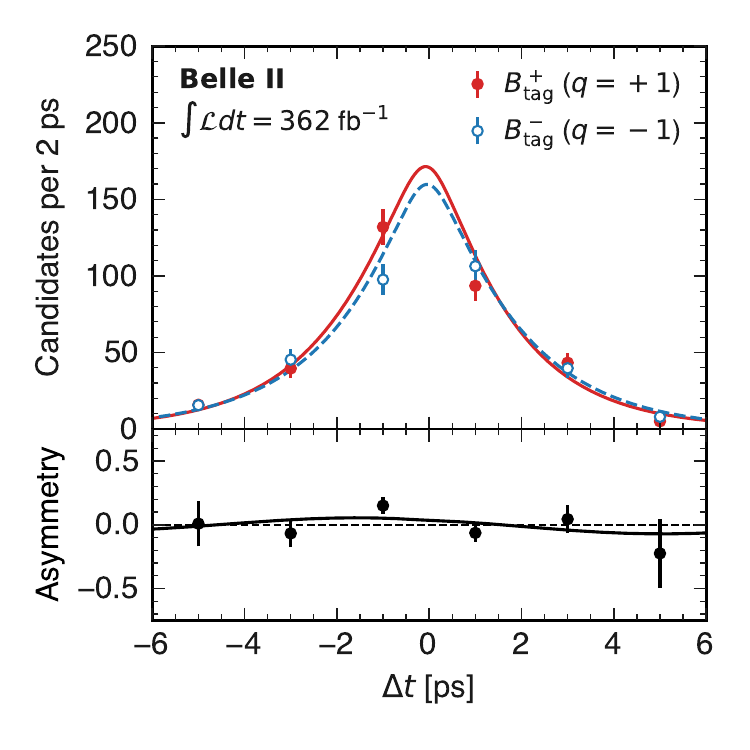}
\caption{Distributions, and fit projections, of \dt for flavor-tagged (left) \phiks and (right) \phikp candidates subtracted of the continuum background.
The fit PDFs corresponding to $q=-1$ and $q=+1$ tagged distributions are shown as dashed and solid curves, respectively.
The yield asymmetries, defined as $(N(q=+1)-N(q=-1))/(N(q=+1)+N(q=-1))$, are displayed in the bottom subpanels.
}
\label{fig:cp-fit}
\end{figure*}

\begin{table}
\centering
\caption{Results of the fit to the signal and control samples.
}
\label{tab:fit-results}
\begin{tabular*}{\columnwidth}{@{\extracolsep{\fill}}lrr@{}}
\hline
\hline
& \phiks & \phikp \\
\hline
Resonant yield    & \ensuremath{162 \pm 17}\xspace & \ensuremath{581 \pm 33}\xspace \\
Nonresonant yield & \ensuremath{21 \pm 12}\xspace & \ensuremath{70 \pm 23}\xspace \\
Continuum   yield & \ensuremath{1169 \pm 35}\xspace  & \ensuremath{5730 \pm 77}\xspace  \\
\hline
\ACP & \cmt{\ensuremath{0.31 \pm 0.20}} & \cmt{\ensuremath{0.12 \pm 0.10}} \\
\SCP & \ensuremath{0.54 \pm 0.26}\xspace & \ensuremath{-0.09 \pm 0.12}\xspace \\
\hline
\hline
\end{tabular*}
\end{table}

The fit results are reported in Table~\ref{tab:fit-results}.
In the control sample, we find \ensuremath{581 \pm 33}\xspace signal \phikp, \ensuremath{70 \pm 23}\xspace nonresonant, and \ensuremath{5730 \pm 77}\xspace continuum events.
The relevant data distributions are displayed in Fig.~\ref{fig:yields-fit-kp}, with fit projections overlaid, under selections in the analysis variables that enhance the signal component.
The control-sample \CP asymmetries are $\ACP=\cmt{\ensuremath{0.12 \pm 0.10}}$ and $\SCP=\ensuremath{-0.09 \pm 0.12}\xspace$, where the uncertainties are statistical only, with correlation coefficient $\rho=\cmt{\ensuremath{-0.06}}\xspace$.
The results are compatible with the null asymmetries we expect.
In the fit to the signal \phiks sample, displayed under the same signal-enhancing selections in Fig.~\ref{fig:yields-fit-ks}, we find \ensuremath{162 \pm 17}\xspace signal, \ensuremath{21 \pm 12}\xspace nonresonant, and \ensuremath{1169 \pm 35}\xspace continuum  events.
The corresponding \CP asymmetries are $\ACP=\cmt{\ensuremath{0.31 \pm 0.20}}$ and $\SCP=\ensuremath{0.54 \pm 0.26}\xspace$, where the uncertainties are statistical only, with correlation coefficient $\rho=\cmt{\ensuremath{-0.01}}\xspace$.
The observed continuum background asymmetry is compatible with zero.
The \dt distributions for tagged signal decays, after subtracting the continuum background~\cite{Pivk:2004ty}, are displayed in Fig.~\ref{fig:cp-fit}, along with the resulting \CP-violating asymmetries.

\section{Systematic uncertainties}
Contributions from all considered sources of systematic uncertainty are listed in Table~\ref{tab:systematics}.
We consider uncertainties associated with the calibration of the flavor tagging and resolution function, fit model, and determination of \dt.

The leading contribution to the total systematic uncertainty on \ACP arises by neglecting a possible time-integrated \CP asymmetry from \BBbar backgrounds.
The main systematic uncertainty on \SCP comes from the fit bias, due to the modest statistical precision to which the fraction of \nonres backgrounds can be determined with the current sample size.

\subsection{Calibration with \dpi decays}

We assess the uncertainty associated with the resolution function and flavor tagging parameters using simplified simulated samples.
We generate ensembles assuming for each an alternative value for the above parameters sampled from the statistical covariance matrix determined in the \dpi control sample.
Each ensemble is fitted using the nominal values of the calibration parameters and the standard deviation of the observed biases is used as a systematic uncertainty.

A similar procedure is used to assess a systematic uncertainty due to the systematic uncertainties on the calibration parameters, in which the ensembles are generated by varying each parameter independently within their systematic uncertainty.

We estimate the impact of differences in the resolution function and tagging performance between the signal and calibration samples.
We apply the resolution function and flavor-tagging calibration obtained from a simulated \dpi sample and repeat the measurement of \ACP and \SCP over an ensemble of simulated \phiks events.
The average deviation of the \CP asymmetries from their generated values is assigned as a systematic uncertainty.

\subsection{Fit model}

To validate how accurately the fit determines the underlying physics parameters in the presence of backgrounds, we generate ensemble datasets that contain all the fit components.
For each ensemble, we sample alternative values of \ACP and \SCP within the physical boundaries, and the fraction of the resonant events over the sum of resonant and nonresonant decays between 0.7 and 1.0, to account for the statistical precision on the observed value $f_{\phi K}=0.89\pm0.07$.
Due to the limited sample size, we assign a conservative systematic uncertainty for the fit bias by taking the largest deviations of the fitted values of \ACP and \SCP from their generated values.
We also check that the relative magnitude of this systematic uncertainty with respect to the statistical uncertainty remains constant for larger sample sizes.

We study the effect of neglecting interference between the signal and nonresonant backgrounds using simulated samples, where the \phiks and \nonres components are generated coherently using a complete Dalitz-plot description of the decay~\cite{PhysRevD.82.073011}.
We apply the nominal fit to these samples, where the nonresonant yields are determined by the fit and the \CP-asymmetries of the backgrounds, \acpnrk and \scpnrk, are fixed to their generated values, neglecting interference with the signal.
The difference between the generated and fitted values of the \CP-asymmetries of the signal is assigned as a systematic uncertainty.

The effect of fixing the PDF shapes of the \mbc, \ocs, \cosh, and \dt distributions in continuum, and \ocs distribution in signal and nonresonant background, is estimated from ensemble datasets.
We generate simulated datasets by varying the shape parameters, in order to cover for the empirical parametrization and statistical uncertainty, and fix them to their nominal values in the fit. 
The resulting standard deviation on the distributions of \ACP and \SCP is used to estimate the corresponding systematic uncertainty.

The same procedure is applied to estimate the systematic uncertainty associated with the external inputs used for the lifetime $\taud=(1.519\pm0.004)~\ps$, mixing frequency $\dmd=(0.507\pm0.002)~\invps$, and \CP asymmetries \acpnrkwa and $\SCP = -0.68^{+0.09}_{-0.10}$ of the nonresonant background.

Simulation shows that the residual \BBbar backgrounds is at most 2\% of the signal yield.
We generate ensemble datasets containing an additional \BBbar background component with PDF shapes modeled after the \phiks or \nonres distributions and by conservatively varying the \BBbar background \CP asymmetries between $+1$ and $-1$.
The \BBbar backgrounds are neglected in the fit to these datasets.
The corresponding systematic uncertainty is obtained by taking the largest deviations of \ACP and \SCP from their generated values.

The time evolution given in Eq.~\eqref{eq:dt_theo} assumes that the \btag decays in a flavor-specific final state.
We study the impact of the tag-side interference, \ie,~neglecting the effect of CKM-suppressed $\bquark \to \uquark\cquarkbar\dquark$ decays in the \btag in the model for \dt~\cite{PhysRevD.68.034010}.
The observed asymmetries can be corrected for this effect by using the knowledge from previous measurements~\cite{HeavyFlavorAveragingGroup:2022wzx}.
We conservatively assume all events to be tagged by hadronic \B decays, for which the effect is largest, and take the difference with respect to the observed asymmetries as a systematic uncertainty.

The effect of multiple candidates is evaluated by repeating the analysis with all the candidates and taking the difference with respect to the nominal candidate selection as a systematic uncertainty.

\begin{table}
\caption{Summary of systematic uncertainties.}
\label{tab:systematics}
\centering
\renewcommand{\arraystretch}{1.5}
\ifbool{useHFLAV}{
\begin{tabular*}{\columnwidth}{@{\extracolsep{\fill}}lrr@{}}
\hline
\hline
Source & $\sigma(\ACP)$ & $\sigma(\SCP)$ \\
\hline
\multicolumn{3}{l}{Calibration with \dpi decays} \\
\quad Calibration sample size &$\pm0.010$ & $\pm0.009$\\
\quad Calibration sample systematic &$\pm0.010$ & $\pm0.012$\\
\quad Sample dependence & $+0.005$ & $+0.021$ \\
\multicolumn{3}{l}{Fit model} \\\quad Fit bias &$ ^{\mbox{\footnotesize$+0.028$}}_{\mbox{\footnotesize$-0.017$}}$ &$ ^{\mbox{\footnotesize$+0.033$}}_{\mbox{\footnotesize$-0.062$}}$ \\
\quad \nonres backgrounds & $+0.020$ & $-0.011$ \\
\quad Fixed fit shapes &$\pm0.009$ & $\pm0.022$\\
\quad \taud and \dmd uncertainties &$\pm0.006$ & $\pm0.022$\\
\quad \acpnrk and \scpnrk &$\pm0.014$ & $\pm0.013$\\
\quad $B\overline{B}$ backgrounds &$ ^{\mbox{\footnotesize$+0.019$}}_{\mbox{\footnotesize$-0.030$}}$ &$ ^{\mbox{\footnotesize$+0.017$}}_{\mbox{\footnotesize$-0.031$}}$ \\
\quad Tag-side interference & $<0.001$ &$+0.012$ \\
\quad Multiple candidates & $-0.032$ & $-0.002$ \\
\multicolumn{3}{l}{\dt measurement} \\\quad Detector misalignment & $-0.002$ & $-0.002$ \\
\quad Momentum scale &$\pm0.001$ & $\pm0.001$\\
\quad Beam spot &$\pm0.002$ & $\pm0.002$\\
\quad $\dt$ approximation &$<0.001$ & $-0.018$ \\
\hline
Total systematic & $^{\mbox{\footnotesize$+0.046$}}_{\mbox{\footnotesize$-0.052$}}$ &$^{\mbox{\footnotesize$+0.058$}}_{\mbox{\footnotesize$-0.082$}}$ \\
\hline
Statistical & $\pm0.201$ & $\pm0.256$\\
\hline
\hline
\end{tabular*}
}{
\begin{tabular*}{\columnwidth}{@{\extracolsep{\fill}}lrr@{}}
\hline
\hline
Source & $\sigma(\ACP)$ & $\sigma(\SCP)$ \\
\hline
\multicolumn{3}{l}{Calibration with \dpi decays} \\
\quad Calibration sample size &$\pm0.010$ & $\pm0.009$\\
\quad Calibration sample systematic &$\pm0.010$ & $\pm0.012$\\
\quad Sample dependence & $-0.005$ & $+0.021$ \\
\multicolumn{3}{l}{Fit model} \\\quad Fit bias &$ ^{\mbox{\footnotesize$+0.017$}}_{\mbox{\footnotesize$-0.028$}}$ &$ ^{\mbox{\footnotesize$+0.033$}}_{\mbox{\footnotesize$-0.062$}}$ \\
\quad \nonres backgrounds & $-0.020$ & $-0.011$ \\
\quad Fixed fit shapes &$\pm0.009$ & $\pm0.022$\\
\quad \taud and \dmd uncertainties &$\pm0.006$ & $\pm0.022$\\
\quad \acpnrk and \scpnrk &$\pm0.014$ & $\pm0.013$\\
\quad $B\overline{B}$ backgrounds &$ ^{\mbox{\footnotesize$+0.030$}}_{\mbox{\footnotesize$-0.019$}}$ &$ ^{\mbox{\footnotesize$+0.017$}}_{\mbox{\footnotesize$-0.031$}}$ \\
\quad Tag-side interference & $<0.001$ &$+0.012$ \\
\quad Multiple candidates & $+0.032$ & $-0.002$ \\
\multicolumn{3}{l}{\dt measurement} \\\quad Detector misalignment & $+0.002$ & $-0.002$ \\
\quad Momentum scale &$\pm0.001$ & $\pm0.001$\\
\quad Beam spot &$\pm0.002$ & $\pm0.002$\\
\quad $\dt$ approximation &$<0.001$ & $-0.018$ \\
\hline
Total systematic & $^{\mbox{\footnotesize$+0.052$}}_{\mbox{\footnotesize$-0.046$}}$ &$^{\mbox{\footnotesize$+0.058$}}_{\mbox{\footnotesize$-0.082$}}$ \\
\hline
Statistical & $\pm0.201$ & $\pm0.256$\\
\hline
\hline
\end{tabular*} 
}
\end{table}

\subsection{\dt measurement}

The impact of the detector misalignment is tested on simulated samples reconstructed with various misalignment configurations.

The uncertainty on the momentum scale of charged particles due to the imperfect modeling of the magnetic field has a small impact on the \CP asymmetries~\cite{Belle-II:2023nmj}. 

Similarly, the uncertainty on the coordinates of the \epem interaction region (beam spot) has a subleading effect~\cite{Belle-II:2023nmj}.

We do not account for the angular distribution of the \B meson pairs in the c.m. frame when calculating \dt using Eq.~\eqref{eq:dt-approx}.
Therefore, we estimate the effect of the \dt approximation on simulated samples, where the generated and reconstructed time differences can be compared.

\section{Summary}
A measurement of \CP violation in \phiks decays is presented using data from the \belletwo experiment.
We find \ensuremath{162 \pm 17}\xspace signal candidates in a sample containing \nbb \BBbar events.
The values of the \CP asymmetries are
\begin{linenomath*}
\begin{align*}
\ACP=\cmt{\ensuremath{0.31 \pm 0.20 \pm0.05}}\xspace\quad \text{and} \quad \SCP=\ensuremath{0.54 \pm 0.26 ^{+0.06}_{-0.08}}\xspace,
\end{align*}
\end{linenomath*}
where the first uncertainty is statistical, and the second is systematic.
The results are compatible with previous determinations from \belle and \babar~\cite{PhysRevD.82.073011,PhysRevD.85.112010} and have a similar uncertainty on \ACP, despite using a data sample $2.0$ and $1.2$ times smaller, respectively.
When compared to measurements using a similar quasi-two-body approach~\cite{BaBar:2005xng,PhysRevLett.98.031802}, there is a 10\% to 20\% improvement on the statistical uncertainty on \SCP for the same number of signal events.
No significant discrepancy in the \CP asymmetries between \qqs and \ccs transitions is observed.

\section*{Acknowledgements}
This work, based on data collected using the Belle II detector, which was built and commissioned prior to March 2019, was supported by
Science Committee of the Republic of Armenia Grant No.~20TTCG-1C010;
Australian Research Council and Research Grants
No.~DP200101792, 
No.~DP210101900, 
No.~DP210102831, 
No.~DE220100462, 
No.~LE210100098, 
and
No.~LE230100085; 
Austrian Federal Ministry of Education, Science and Research,
Austrian Science Fund
No.~P~31361-N36
and
No.~J4625-N,
and
Horizon 2020 ERC Starting Grant No.~947006 ``InterLeptons'';
Natural Sciences and Engineering Research Council of Canada, Compute Canada and CANARIE;
National Key R\&D Program of China under Contract No.~2022YFA1601903,
National Natural Science Foundation of China and Research Grants
No.~11575017,
No.~11761141009,
No.~11705209,
No.~11975076,
No.~12135005,
No.~12150004,
No.~12161141008,
and
No.~12175041,
and Shandong Provincial Natural Science Foundation Project~ZR2022JQ02;
the Ministry of Education, Youth, and Sports of the Czech Republic under Contract No.~LTT17020 and
Charles University Grant No.~SVV 260448 and
the Czech Science Foundation Grant No.~22-18469S;
European Research Council, Seventh Framework PIEF-GA-2013-622527,
Horizon 2020 ERC-Advanced Grants No.~267104 and No.~884719,
Horizon 2020 ERC-Consolidator Grant No.~819127,
Horizon 2020 Marie Sklodowska-Curie Grant Agreement No.~700525 ``NIOBE''
and
No.~101026516,
and
Horizon 2020 Marie Sklodowska-Curie RISE project JENNIFER2 Grant Agreement No.~822070 (European grants);
L'Institut National de Physique Nucl\'{e}aire et de Physique des Particules (IN2P3) du CNRS (France);
BMBF, DFG, HGF, MPG, and AvH Foundation (Germany);
Department of Atomic Energy under Project Identification No.~RTI 4002 and Department of Science and Technology (India);
Israel Science Foundation Grant No.~2476/17,
U.S.-Israel Binational Science Foundation Grant No.~2016113, and
Israel Ministry of Science Grant No.~3-16543;
Istituto Nazionale di Fisica Nucleare and the Research Grants BELLE2;
Japan Society for the Promotion of Science, Grant-in-Aid for Scientific Research Grants
No.~16H03968,
No.~16H03993,
No.~16H06492,
No.~16K05323,
No.~17H01133,
No.~17H05405,
No.~18K03621,
No.~18H03710,
No.~18H05226,
No.~19H00682, 
No.~22H00144,
No.~26220706,
and
No.~26400255,
the National Institute of Informatics, and Science Information NETwork 5 (SINET5), 
and
the Ministry of Education, Culture, Sports, Science, and Technology (MEXT) of Japan;  
National Research Foundation (NRF) of Korea Grants
No.~2016R1\-D1A1B\-02012900,
No.~2018R1\-A2B\-3003643,
No.~2018R1\-A6A1A\-06024970,
No.~2019R1\-I1A3A\-01058933,
No.~2021R1\-A6A1A\-03043957,
No.~2021R1\-F1A\-1060423,
No.~2021R1\-F1A\-1064008,
No.~2022R1\-A2C\-1003993,
and
No.~RS-2022-00197659,
Radiation Science Research Institute,
Foreign Large-size Research Facility Application Supporting project,
the Global Science Experimental Data Hub Center of the Korea Institute of Science and Technology Information
and
KREONET/GLORIAD;
Universiti Malaya RU grant, Akademi Sains Malaysia, and Ministry of Education Malaysia;
Frontiers of Science Program Contracts
No.~FOINS-296,
No.~CB-221329,
No.~CB-236394,
No.~CB-254409,
and
No.~CB-180023, and SEP-CINVESTAV Research Grant No.~237 (Mexico);
the Polish Ministry of Science and Higher Education and the National Science Center;
the Ministry of Science and Higher Education of the Russian Federation,
Agreement No.~14.W03.31.0026, and
the HSE University Basic Research Program, Moscow;
University of Tabuk Research Grants
No.~S-0256-1438 and No.~S-0280-1439 (Saudi Arabia);
Slovenian Research Agency and Research Grants
No.~J1-9124
and
No.~P1-0135;
Agencia Estatal de Investigacion, Spain
Grant No.~RYC2020-029875-I
and
Generalitat Valenciana, Spain
Grant No.~CIDEGENT/2018/020
Ministry of Science and Technology and Research Grants
No.~MOST106-2112-M-002-005-MY3
and
No.~MOST107-2119-M-002-035-MY3,
and the Ministry of Education (Taiwan);
Thailand Center of Excellence in Physics;
TUBITAK ULAKBIM (Turkey);
National Research Foundation of Ukraine, Project No.~2020.02/0257,
and
Ministry of Education and Science of Ukraine;
the U.S. National Science Foundation and Research Grants
No.~PHY-1913789 
and
No.~PHY-2111604, 
and the U.S. Department of Energy and Research Awards
No.~DE-AC06-76RLO1830, 
No.~DE-SC0007983, 
No.~DE-SC0009824, 
No.~DE-SC0009973, 
No.~DE-SC0010007, 
No.~DE-SC0010073, 
No.~DE-SC0010118, 
No.~DE-SC0010504, 
No.~DE-SC0011784, 
No.~DE-SC0012704, 
No.~DE-SC0019230, 
No.~DE-SC0021274, 
No.~DE-SC0022350, 
No.~DE-SC0023470; 
and
the Vietnam Academy of Science and Technology (VAST) under Grant No.~DL0000.05/21-23.

These acknowledgements are not to be interpreted as an endorsement of any statement made
by any of our institutes, funding agencies, governments, or their representatives.

We thank the SuperKEKB team for delivering high-luminosity collisions;
the KEK cryogenics group for the efficient operation of the detector solenoid magnet;
the KEK computer group and the NII for on-site computing support and SINET6 network support;
and the raw-data centers at BNL, DESY, GridKa, IN2P3, INFN, and the University of Victoria for offsite computing support.

\bibliographystyle{apsrev4-1}
\bibliography{references}

\begin{thebibliography}{35}%
\makeatletter
\providecommand \@ifxundefined [1]{%
 \@ifx{#1\undefined}
}%
\providecommand \@ifnum [1]{%
 \ifnum #1\expandafter \@firstoftwo
 \else \expandafter \@secondoftwo
 \fi
}%
\providecommand \@ifx [1]{%
 \ifx #1\expandafter \@firstoftwo
 \else \expandafter \@secondoftwo
 \fi
}%
\providecommand \natexlab [1]{#1}%
\providecommand \enquote  [1]{``#1''}%
\providecommand \bibnamefont  [1]{#1}%
\providecommand \bibfnamefont [1]{#1}%
\providecommand \citenamefont [1]{#1}%
\providecommand \href@noop [0]{\@secondoftwo}%
\providecommand \href [0]{\begingroup \@sanitize@url \@href}%
\providecommand \@href[1]{\@@startlink{#1}\@@href}%
\providecommand \@@href[1]{\endgroup#1\@@endlink}%
\providecommand \@sanitize@url [0]{\catcode `\\12\catcode `\$12\catcode
  `\&12\catcode `\#12\catcode `\^12\catcode `\_12\catcode `\%12\relax}%
\providecommand \@@startlink[1]{}%
\providecommand \@@endlink[0]{}%
\providecommand \url  [0]{\begingroup\@sanitize@url \@url }%
\providecommand \@url [1]{\endgroup\@href {#1}{\urlprefix }}%
\providecommand \urlprefix  [0]{URL }%
\providecommand \Eprint [0]{\href }%
\providecommand \doibase [0]{http://dx.doi.org/}%
\providecommand \selectlanguage [0]{\@gobble}%
\providecommand \bibinfo  [0]{\@secondoftwo}%
\providecommand \bibfield  [0]{\@secondoftwo}%
\providecommand \translation [1]{[#1]}%
\providecommand \BibitemOpen [0]{}%
\providecommand \bibitemStop [0]{}%
\providecommand \bibitemNoStop [0]{.\EOS\space}%
\providecommand \EOS [0]{\spacefactor3000\relax}%
\providecommand \BibitemShut  [1]{\csname bibitem#1\endcsname}%
\let\auto@bib@innerbib\@empty
\bibitem [{\citenamefont {Cabibbo}(1963)}]{Cabibbo:1963yz}%
  \BibitemOpen
  \bibfield  {author} {\bibinfo {author} {\bibfnamefont {N.}~\bibnamefont
  {Cabibbo}},\ }\href {\doibase 10.1103/PhysRevLett.10.531} {\bibfield
  {journal} {\bibinfo  {journal} {Phys. Rev. Lett.}\ }\textbf {\bibinfo
  {volume} {10}},\ \bibinfo {pages} {531} (\bibinfo {year} {1963})}\BibitemShut
  {NoStop}%
\bibitem [{\citenamefont {Kobayashi}\ and\ \citenamefont
  {Maskawa}(1973)}]{Kobayashi:1973fv}%
  \BibitemOpen
  \bibfield  {author} {\bibinfo {author} {\bibfnamefont {M.}~\bibnamefont
  {Kobayashi}}\ and\ \bibinfo {author} {\bibfnamefont {T.}~\bibnamefont
  {Maskawa}},\ }\href {\doibase 10.1143/PTP.49.652} {\bibfield  {journal}
  {\bibinfo  {journal} {Prog. Theor. Phys.}\ }\textbf {\bibinfo {volume}
  {49}},\ \bibinfo {pages} {652} (\bibinfo {year} {1973})}\BibitemShut
  {NoStop}%
\bibitem [{\citenamefont {Amhis}\ \emph {et~al.}(2023)\citenamefont {Amhis}
  \emph {et~al.}}]{HeavyFlavorAveragingGroup:2022wzx}%
  \BibitemOpen
  \bibfield  {author} {\bibinfo {author} {\bibfnamefont {Y.~S.}\ \bibnamefont
  {Amhis}} \emph {et~al.} (\bibinfo {collaboration} {{HFLAV Collaboration}}),\
  }\href {\doibase https://doi.org/10.1103/PhysRevD.107.052008} {\bibfield
  {journal} {\bibinfo  {journal} {Phys. Rev. D}\ }\textbf {\bibinfo {volume}
  {107}},\ \bibinfo {pages} {052008} (\bibinfo {year} {2023})}\BibitemShut
  {NoStop}%
\bibitem [{\citenamefont {Beneke}(2005)}]{Beneke:2005pu}%
  \BibitemOpen
  \bibfield  {author} {\bibinfo {author} {\bibfnamefont {M.}~\bibnamefont
  {Beneke}},\ }\href {\doibase 10.1016/j.physletb.2005.06.045} {\bibfield
  {journal} {\bibinfo  {journal} {Phys. Lett. B}\ }\textbf {\bibinfo {volume}
  {620}},\ \bibinfo {pages} {143} (\bibinfo {year} {2005})}\BibitemShut
  {NoStop}%
\bibitem [{cp-()}]{cp-coeffs}%
  \BibitemOpen
  \href@noop {} {}\bibinfo {howpublished} {{The coefficients ($S$,$-C$) are
  written ($S$, $A$) elsewhere.}}\BibitemShut {Stop}%
\bibitem [{\citenamefont {Workman}\ \emph {et~al.}(2022)\citenamefont {Workman}
  \emph {et~al.}}]{Workman:2022ynf}%
  \BibitemOpen
  \bibfield  {author} {\bibinfo {author} {\bibfnamefont {R.~L.}\ \bibnamefont
  {Workman}} \emph {et~al.} (\bibinfo {collaboration} {Particle Data Group}),\
  }\href {\doibase 10.1093/ptep/ptac097} {\bibfield  {journal} {\bibinfo
  {journal} {Prog. Theor. Exp. Phys.}\ }\textbf {\bibinfo {volume} {2022}},\
  \bibinfo {pages} {083C01} (\bibinfo {year} {2022})}\BibitemShut {NoStop}%
\bibitem [{\citenamefont {Abudin\'en}\ \emph {et~al.}(2022)\citenamefont
  {Abudin\'en} \emph {et~al.}}]{Belle-II:2021zvj}%
  \BibitemOpen
  \bibfield  {author} {\bibinfo {author} {\bibfnamefont {F.}~\bibnamefont
  {Abudin\'en}} \emph {et~al.} (\bibinfo {collaboration} {{Belle II
  Collaboration}}),\ }\href {\doibase 10.1140/epjc/s10052-022-10180-9}
  {\bibfield  {journal} {\bibinfo  {journal} {Eur. Phys. J. C}\ }\textbf
  {\bibinfo {volume} {82}},\ \bibinfo {pages} {283} (\bibinfo {year}
  {2022})}\BibitemShut {NoStop}%
\bibitem [{\citenamefont {Nakahama}\ \emph {et~al.}(2010)\citenamefont
  {Nakahama} \emph {et~al.}}]{PhysRevD.82.073011}%
  \BibitemOpen
  \bibfield  {author} {\bibinfo {author} {\bibfnamefont {Y.}~\bibnamefont
  {Nakahama}} \emph {et~al.} (\bibinfo {collaboration} {Belle Collaboration}),\
  }\href {\doibase 10.1103/PhysRevD.82.073011} {\bibfield  {journal} {\bibinfo
  {journal} {Phys. Rev. D}\ }\textbf {\bibinfo {volume} {82}},\ \bibinfo
  {pages} {073011} (\bibinfo {year} {2010})}\BibitemShut {NoStop}%
\bibitem [{\citenamefont {Lees}\ \emph {et~al.}(2012)\citenamefont {Lees} \emph
  {et~al.}}]{PhysRevD.85.112010}%
  \BibitemOpen
  \bibfield  {author} {\bibinfo {author} {\bibfnamefont {J.~P.}\ \bibnamefont
  {Lees}} \emph {et~al.} (\bibinfo {collaboration} {\babar Collaboration}),\
  }\href {\doibase 10.1103/PhysRevD.85.112010} {\bibfield  {journal} {\bibinfo
  {journal} {Phys. Rev. D}\ }\textbf {\bibinfo {volume} {85}},\ \bibinfo
  {pages} {112010} (\bibinfo {year} {2012})}\BibitemShut {NoStop}%
\bibitem [{\citenamefont {Abe}\ \emph {et~al.}()\citenamefont {Abe} \emph
  {et~al.}}]{Abe:2010gxa}%
  \BibitemOpen
  \bibfield  {author} {\bibinfo {author} {\bibfnamefont {T.}~\bibnamefont
  {Abe}} \emph {et~al.} (\bibinfo {collaboration} {Belle II Collaboration}),\
  }\href@noop {} {\ }\Eprint {http://arxiv.org/abs/1011.0352} {arXiv:1011.0352}
  \BibitemShut {NoStop}%
\bibitem [{\citenamefont {Adamczyk}\ \emph {et~al.}(2022)\citenamefont
  {Adamczyk} \emph {et~al.}}]{Belle-IISVD:2022upf}%
  \BibitemOpen
  \bibfield  {author} {\bibinfo {author} {\bibfnamefont {K.}~\bibnamefont
  {Adamczyk}} \emph {et~al.} (\bibinfo {collaboration} {Belle II SVD
  Collaboration}),\ }\href {\doibase 10.1088/1748-0221/17/11/P11042} {\bibfield
   {journal} {\bibinfo  {journal} {J. Instrum.}\ }\textbf {\bibinfo {volume}
  {17}},\ \bibinfo {pages} {P11042} (\bibinfo {year} {2022})}\BibitemShut
  {NoStop}%
\bibitem [{\citenamefont {Jadach}\ \emph {et~al.}(2000)\citenamefont {Jadach},
  \citenamefont {Ward},\ and\ \citenamefont {W\c{a}s}}]{Jadach:1999vf}%
  \BibitemOpen
  \bibfield  {author} {\bibinfo {author} {\bibfnamefont {S.}~\bibnamefont
  {Jadach}}, \bibinfo {author} {\bibfnamefont {B.~F.~L.}\ \bibnamefont {Ward}},
  \ and\ \bibinfo {author} {\bibfnamefont {Z.}~\bibnamefont {W\c{a}s}},\ }\href
  {\doibase 10.1016/S0010-4655(00)00048-5} {\bibfield  {journal} {\bibinfo
  {journal} {Comput. Phys. Commun.}\ }\textbf {\bibinfo {volume} {130}},\
  \bibinfo {pages} {260} (\bibinfo {year} {2000})}\BibitemShut {NoStop}%
\bibitem [{\citenamefont {Sj\"{o}strand}\ \emph {et~al.}(2015)\citenamefont
  {Sj\"{o}strand} \emph {et~al.}}]{Sjostrand:2014zea}%
  \BibitemOpen
  \bibfield  {author} {\bibinfo {author} {\bibfnamefont {T.}~\bibnamefont
  {Sj\"{o}strand}} \emph {et~al.},\ }\href {\doibase 10.1016/j.cpc.2015.01.024}
  {\bibfield  {journal} {\bibinfo  {journal} {Comput. Phys. Commun.}\ }\textbf
  {\bibinfo {volume} {191}},\ \bibinfo {pages} {159} (\bibinfo {year}
  {2015})}\BibitemShut {NoStop}%
\bibitem [{\citenamefont {Lange}(2001)}]{Lange:2001uf}%
  \BibitemOpen
  \bibfield  {author} {\bibinfo {author} {\bibfnamefont {D.~J.}\ \bibnamefont
  {Lange}},\ }\href {\doibase 10.1016/S0168-9002(01)00089-4} {\bibfield
  {journal} {\bibinfo  {journal} {Nucl. Instrum. Methods Phys. Res., Sect A}\
  }\textbf {\bibinfo {volume} {{462}}},\ \bibinfo {pages} {152} (\bibinfo
  {year} {2001})}\BibitemShut {NoStop}%
\bibitem [{\citenamefont {Agostinelli}\ \emph {et~al.}(2003)\citenamefont
  {Agostinelli} \emph {et~al.}}]{Agostinelli:2002hh}%
  \BibitemOpen
  \bibfield  {author} {\bibinfo {author} {\bibfnamefont {S.}~\bibnamefont
  {Agostinelli}} \emph {et~al.} (\bibinfo {collaboration} {GEANT4
  Collaboration}),\ }\href {\doibase 10.1016/S0168-9002(03)01368-8} {\bibfield
  {journal} {\bibinfo  {journal} {Nucl. Instrum. Methods Phys. Res., Sect. A}\
  }\textbf {\bibinfo {volume} {{506}}},\ \bibinfo {pages} {250} (\bibinfo
  {year} {2003})}\BibitemShut {NoStop}%
\bibitem [{\citenamefont {Kuhr}\ \emph {et~al.}(2019)\citenamefont {Kuhr},
  \citenamefont {Pulvermacher}, \citenamefont {Ritter}, \citenamefont {Hauth},\
  and\ \citenamefont {Braun}}]{Kuhr:2018lps}%
  \BibitemOpen
  \bibfield  {author} {\bibinfo {author} {\bibfnamefont {T.}~\bibnamefont
  {Kuhr}}, \bibinfo {author} {\bibfnamefont {C.}~\bibnamefont {Pulvermacher}},
  \bibinfo {author} {\bibfnamefont {M.}~\bibnamefont {Ritter}}, \bibinfo
  {author} {\bibfnamefont {T.}~\bibnamefont {Hauth}}, \ and\ \bibinfo {author}
  {\bibfnamefont {N.}~\bibnamefont {Braun}} (\bibinfo {collaboration} {Belle II
  Framework Software Group}),\ }\href {\doibase 10.1007/s41781-018-0017-9}
  {\bibfield  {journal} {\bibinfo  {journal} {Comput. Software Big Sci.}\
  }\textbf {\bibinfo {volume} {3}},\ \bibinfo {pages} {1} (\bibinfo {year}
  {2019})}\BibitemShut {NoStop}%
\bibitem [{bas()}]{basf2-zenodo}%
  \BibitemOpen
  \href@noop {} {}\bibinfo {howpublished}
  {\url{https://doi.org/10.5281/zenodo.5574115}}\BibitemShut {NoStop}%
\bibitem [{\citenamefont {Bertacchi}\ \emph {et~al.}(2021)\citenamefont
  {Bertacchi} \emph {et~al.}}]{Bertacchi:2020eez}%
  \BibitemOpen
  \bibfield  {author} {\bibinfo {author} {\bibfnamefont {V.}~\bibnamefont
  {Bertacchi}} \emph {et~al.} (\bibinfo {collaboration} {Belle II Tracking
  Group}),\ }\href {\doibase 10.1016/j.cpc.2020.107610} {\bibfield  {journal}
  {\bibinfo  {journal} {Comput. Phys. Commun.}\ }\textbf {\bibinfo {volume}
  {259}},\ \bibinfo {pages} {107610} (\bibinfo {year} {2021})}\BibitemShut
  {NoStop}%
\bibitem [{\citenamefont {Hulsbergen}(2005)}]{HULSBERGEN2005566}%
  \BibitemOpen
  \bibfield  {author} {\bibinfo {author} {\bibfnamefont {W.~D.}\ \bibnamefont
  {Hulsbergen}},\ }\href {\doibase https://doi.org/10.1016/j.nima.2005.06.078}
  {\bibfield  {journal} {\bibinfo  {journal} {Nucl. Instrum. Methods}\ }\textbf
  {\bibinfo {volume} {552}},\ \bibinfo {pages} {566} (\bibinfo {year}
  {2005})}\BibitemShut {NoStop}%
\bibitem [{\citenamefont {Krohn}\ \emph {et~al.}(2020)\citenamefont {Krohn}
  \emph {et~al.}}]{Krohn:2019dlq}%
  \BibitemOpen
  \bibfield  {author} {\bibinfo {author} {\bibfnamefont {J.-F.}\ \bibnamefont
  {Krohn}} \emph {et~al.} (\bibinfo {collaboration} {Belle II Analysis Software
  Group}),\ }\href {\doibase 10.1016/j.nima.2020.164269} {\bibfield  {journal}
  {\bibinfo  {journal} {Nucl. Instrum. Methods Phys. Res., Sect. A}\ }\textbf
  {\bibinfo {volume} {{976}}},\ \bibinfo {pages} {164269} (\bibinfo {year}
  {2020})}\BibitemShut {NoStop}%
\bibitem [{\citenamefont {Waltenberger}\ \emph {et~al.}(2008)\citenamefont
  {Waltenberger}, \citenamefont {Mitaroff}, \citenamefont {Moser},
  \citenamefont {Pflugfelder},\ and\ \citenamefont
  {Riedel}}]{Waltenberger_2008}%
  \BibitemOpen
  \bibfield  {author} {\bibinfo {author} {\bibfnamefont {W.}~\bibnamefont
  {Waltenberger}}, \bibinfo {author} {\bibfnamefont {W.}~\bibnamefont
  {Mitaroff}}, \bibinfo {author} {\bibfnamefont {F.}~\bibnamefont {Moser}},
  \bibinfo {author} {\bibfnamefont {B.}~\bibnamefont {Pflugfelder}}, \ and\
  \bibinfo {author} {\bibfnamefont {H.~V.}\ \bibnamefont {Riedel}},\ }\href
  {\doibase 10.1088/1742-6596/119/3/032037} {\bibfield  {journal} {\bibinfo
  {journal} {J. Phys. Conf. Ser.}\ }\textbf {\bibinfo {volume} {119}},\
  \bibinfo {pages} {032037} (\bibinfo {year} {2008})}\BibitemShut {NoStop}%
\bibitem [{\citenamefont {Dey}\ and\ \citenamefont
  {Soffer}(2020)}]{btube-conf}%
  \BibitemOpen
  \bibfield  {author} {\bibinfo {author} {\bibfnamefont {S.}~\bibnamefont
  {Dey}}\ and\ \bibinfo {author} {\bibfnamefont {A.}~\bibnamefont {Soffer}},\
  }\href@noop {} {\bibfield  {journal} {\bibinfo  {journal} {Springer Proc.
  Phys.}\ }\textbf {\bibinfo {volume} {248}},\ \bibinfo {pages} {411} (\bibinfo
  {year} {2020})}\BibitemShut {NoStop}%
\bibitem [{\citenamefont {Chen}\ and\ \citenamefont
  {Guestrin}(2016)}]{DBLP:journals/corr/ChenG16}%
  \BibitemOpen
  \bibfield  {author} {\bibinfo {author} {\bibfnamefont {T.}~\bibnamefont
  {Chen}}\ and\ \bibinfo {author} {\bibfnamefont {C.}~\bibnamefont
  {Guestrin}},\ }\bibfield  {booktitle} {\emph {\bibinfo {booktitle}
  {{Proceedings of the 22nd {ACM} {SIGKDD} International Conference on
  Knowledge Discovery and Data Mining}}},\ }\href@noop {} {\  (\bibinfo {year}
  {2016})}\BibitemShut {NoStop}%
\bibitem [{\citenamefont {{Ed.~A.~J.~Bevan, B.~Golob, Th.~Mannel, S.~Prell, and
  B.~D.~Yabsley}}(2014)}]{BaBar:2014omp}%
  \BibitemOpen
  \bibfield  {author} {\bibinfo {author} {\bibnamefont {{Ed.~A.~J.~Bevan,
  B.~Golob, Th.~Mannel, S.~Prell, and B.~D.~Yabsley}}},\ }\href {\doibase
  10.1140/epjc/s10052-014-3026-9} {\bibfield  {journal} {\bibinfo  {journal}
  {Eur. Phys. J. C}\ }\textbf {\bibinfo {volume} {74}},\ \bibinfo {pages}
  {3026} (\bibinfo {year} {2014})}\BibitemShut {NoStop}%
\bibitem [{\citenamefont {Lee}\ \emph {et~al.}(2003)\citenamefont {Lee} \emph
  {et~al.}}]{PhysRevLett.91.261801}%
  \BibitemOpen
  \bibfield  {author} {\bibinfo {author} {\bibfnamefont {S.~H.}\ \bibnamefont
  {Lee}} \emph {et~al.} (\bibinfo {collaboration} {Belle Collaboration}),\
  }\href {\doibase 10.1103/PhysRevLett.91.261801} {\bibfield  {journal}
  {\bibinfo  {journal} {Phys. Rev. Lett.}\ }\textbf {\bibinfo {volume} {91}},\
  \bibinfo {pages} {261801} (\bibinfo {year} {2003})}\BibitemShut {NoStop}%
\bibitem [{\citenamefont {Brandt}\ \emph {et~al.}(1964)\citenamefont {Brandt},
  \citenamefont {Peyrou}, \citenamefont {Sosnowski},\ and\ \citenamefont
  {Wroblewski}}]{Brandt:1964sa}%
  \BibitemOpen
  \bibfield  {author} {\bibinfo {author} {\bibfnamefont {S.}~\bibnamefont
  {Brandt}}, \bibinfo {author} {\bibfnamefont {C.}~\bibnamefont {Peyrou}},
  \bibinfo {author} {\bibfnamefont {R.}~\bibnamefont {Sosnowski}}, \ and\
  \bibinfo {author} {\bibfnamefont {A.}~\bibnamefont {Wroblewski}},\ }\href
  {\doibase 10.1016/0031-9163(64)91176-X} {\bibfield  {journal} {\bibinfo
  {journal} {Phys. Lett.}\ }\textbf {\bibinfo {volume} {12}},\ \bibinfo {pages}
  {57} (\bibinfo {year} {1964})}\BibitemShut {NoStop}%
\bibitem [{\citenamefont {Farhi}(1977)}]{Farhi:1977sg}%
  \BibitemOpen
  \bibfield  {author} {\bibinfo {author} {\bibfnamefont {E.}~\bibnamefont
  {Farhi}},\ }\href {\doibase 10.1103/PhysRevLett.39.1587} {\bibfield
  {journal} {\bibinfo  {journal} {Phys. Rev. Lett.}\ }\textbf {\bibinfo
  {volume} {39}},\ \bibinfo {pages} {1587} (\bibinfo {year}
  {1977})}\BibitemShut {NoStop}%
\bibitem [{\citenamefont {Fox}\ and\ \citenamefont
  {Wolfram}(1978)}]{PhysRevLett.41.1581}%
  \BibitemOpen
  \bibfield  {author} {\bibinfo {author} {\bibfnamefont {G.~C.}\ \bibnamefont
  {Fox}}\ and\ \bibinfo {author} {\bibfnamefont {S.}~\bibnamefont {Wolfram}},\
  }\href {\doibase 10.1103/PhysRevLett.41.1581} {\bibfield  {journal} {\bibinfo
   {journal} {Phys. Rev. Lett.}\ }\textbf {\bibinfo {volume} {41}},\ \bibinfo
  {pages} {1581} (\bibinfo {year} {1978})}\BibitemShut {NoStop}%
\bibitem [{\citenamefont {Albrecht}\ \emph {et~al.}(1990)\citenamefont
  {Albrecht} \emph {et~al.}}]{ALBRECHT1990278}%
  \BibitemOpen
  \bibfield  {author} {\bibinfo {author} {\bibfnamefont {H.}~\bibnamefont
  {Albrecht}} \emph {et~al.} (\bibinfo {collaboration} {ARGUS Collaboration}),\
  }\href {\doibase https://doi.org/10.1016/0370-2693(90)91293-K} {\bibfield
  {journal} {\bibinfo  {journal} {Phys. Lett. B}\ }\textbf {\bibinfo {volume}
  {241}},\ \bibinfo {pages} {278} (\bibinfo {year} {1990})}\BibitemShut
  {NoStop}%
\bibitem [{\citenamefont {Abudin\'en}\ \emph {et~al.}(2023)\citenamefont
  {Abudin\'en} \emph {et~al.}}]{Belle-II:2023bps}%
  \BibitemOpen
  \bibfield  {author} {\bibinfo {author} {\bibfnamefont {F.}~\bibnamefont
  {Abudin\'en}} \emph {et~al.} (\bibinfo {collaboration} {\belletwo
  Collaboration}),\ }\href {\doibase 10.1103/PhysRevD.107.L091102} {\bibfield
  {journal} {\bibinfo  {journal} {Phys. Rev. D}\ }\textbf {\bibinfo {volume}
  {107}},\ \bibinfo {pages} {L091102} (\bibinfo {year} {2023})}\BibitemShut
  {NoStop}%
\bibitem [{\citenamefont {Adachi}\ \emph {et~al.}()\citenamefont {Adachi} \emph
  {et~al.}}]{Belle-II:2023nmj}%
  \BibitemOpen
  \bibfield  {author} {\bibinfo {author} {\bibfnamefont {I.}~\bibnamefont
  {Adachi}} \emph {et~al.} (\bibinfo {collaboration} {{Belle II
  Collaboration}}),\ }\href@noop {} {\ }\Eprint
  {http://arxiv.org/abs/2302.12898} {arXiv:2302.12898} \BibitemShut {NoStop}%
\bibitem [{\citenamefont {Pivk}\ and\ \citenamefont
  {Le~Diberder}(2005)}]{Pivk:2004ty}%
  \BibitemOpen
  \bibfield  {author} {\bibinfo {author} {\bibfnamefont {M.}~\bibnamefont
  {Pivk}}\ and\ \bibinfo {author} {\bibfnamefont {F.~R.}\ \bibnamefont
  {Le~Diberder}},\ }\href {\doibase 10.1016/j.nima.2005.08.106} {\bibfield
  {journal} {\bibinfo  {journal} {Nucl. Instrum. Methods Phys. Res., Sect. A}\
  }\textbf {\bibinfo {volume} {{555}}},\ \bibinfo {pages} {356} (\bibinfo
  {year} {2005})}\BibitemShut {NoStop}%
\bibitem [{\citenamefont {Long}\ \emph {et~al.}(2003)\citenamefont {Long},
  \citenamefont {Baak}, \citenamefont {Cahn},\ and\ \citenamefont
  {Kirkby}}]{PhysRevD.68.034010}%
  \BibitemOpen
  \bibfield  {author} {\bibinfo {author} {\bibfnamefont {O.}~\bibnamefont
  {Long}}, \bibinfo {author} {\bibfnamefont {M.}~\bibnamefont {Baak}}, \bibinfo
  {author} {\bibfnamefont {R.~N.}\ \bibnamefont {Cahn}}, \ and\ \bibinfo
  {author} {\bibfnamefont {D.}~\bibnamefont {Kirkby}},\ }\href {\doibase
  10.1103/PhysRevD.68.034010} {\bibfield  {journal} {\bibinfo  {journal} {Phys.
  Rev. D}\ }\textbf {\bibinfo {volume} {68}},\ \bibinfo {pages} {034010}
  (\bibinfo {year} {2003})}\BibitemShut {NoStop}%
\bibitem [{\citenamefont {Aubert}\ \emph {et~al.}(2005)\citenamefont {Aubert}
  \emph {et~al.}}]{BaBar:2005xng}%
  \BibitemOpen
  \bibfield  {author} {\bibinfo {author} {\bibfnamefont {B.}~\bibnamefont
  {Aubert}} \emph {et~al.} (\bibinfo {collaboration} {\babar Collaboration}),\
  }\href {\doibase 10.1103/PhysRevD.71.091102} {\bibfield  {journal} {\bibinfo
  {journal} {Phys. Rev. D}\ }\textbf {\bibinfo {volume} {71}},\ \bibinfo
  {pages} {091102} (\bibinfo {year} {2005})}\BibitemShut {NoStop}%
\bibitem [{\citenamefont {Chen}\ \emph {et~al.}(2007)\citenamefont {Chen} \emph
  {et~al.}}]{PhysRevLett.98.031802}%
  \BibitemOpen
  \bibfield  {author} {\bibinfo {author} {\bibfnamefont {K.-F.}\ \bibnamefont
  {Chen}} \emph {et~al.} (\bibinfo {collaboration} {Belle Collaboration}),\
  }\href {\doibase 10.1103/PhysRevLett.98.031802} {\bibfield  {journal}
  {\bibinfo  {journal} {Phys. Rev. Lett.}\ }\textbf {\bibinfo {volume} {98}},\
  \bibinfo {pages} {031802} (\bibinfo {year} {2007})}\BibitemShut {NoStop}%
\end{thebibliography}%

\end{document}